\documentclass[twocolumn,showpacs,nofootinbib,preprintnumbers,amsmath,amssymb,showkeys,superscriptaddress]{revtex4-1}

\usepackage[pdftex]{graphicx}  
\usepackage{amsmath}
\usepackage{color}
\usepackage{slashed}
\usepackage[normalem]{ulem}
\usepackage{dsfont}
\usepackage{subcaption}
\usepackage[font=small,justification=justified,format=plain]{caption}
\usepackage{xcolor}
\usepackage{enumitem}
\usepackage{makecell}
\usepackage{multirow}
\usepackage{hyperref}
\hypersetup{colorlinks=true,linkcolor=blue,filecolor=magenta,urlcolor=blue,citecolor=blue}
\usepackage[capitalise]{cleveref}

\newcommand{\blue}[1]{\textcolor{blue}{#1}}

\begin{document}

\title{Nonrelativistic meson masses from the Curci-Ferrari model}

\author{Anaclara Alvez}%
\affiliation{Laboratoire Interdisciplinaire des Sciences du Numérique (LISN),
Paris-Saclay University, 91190, Gif-sur-Yvette, France.}%
\author{Nahuel Barrios}%
\affiliation{%
Instituto de F\'{\i}sica, Facultad de Ingenier\'{\i}a, Universidad de
la Rep\'ublica, J. H. y Reissig 565, 11000 Montevideo, Uruguay.
}%
\author{Florencia Ben\'itez}%
\affiliation{%
Instituto de F\'{\i}sica, Facultad de Ingenier\'{\i}a, Universidad de
la Rep\'ublica, J. H. y Reissig 565, 11000 Montevideo, Uruguay.
}%
\author{Marcela Pel\'aez}%
\affiliation{%
 Instituto de F\'{\i}sica, Facultad de Ingenier\'{\i}a, Universidad de
 la Rep\'ublica, J. H. y Reissig 565, 11000 Montevideo, Uruguay.
}%

\date{\today}

\begin{abstract}
We study the mass spectrum of nonrelativistic mesons composed of charm and bottom quarks within the framework of the Curci-Ferrari model in the Landau gauge, focusing on the influence of the gluon mass on our results. We derive the Hamiltonian from the scattering amplitude of a single massive gluon exchange. By incorporating a confining Cornell potential we solve the Schrödinger equation for the dominant terms of the Hamiltonian, which include the kinetic energy and a Yukawa-type potential. Corrections to the energy are then introduced perturbatively. By studying the parameter space of the model we fit the experimental mass spectrum of charmonium, bottomonium, and charm-bottom mesons. From the five parameters of our approach, we allow the gluon mass and the gauge coupling to run with the energy scale according to the ultraviolet one-loop renormalization flows, computed in \cite{Pelaez:2014mxa}. Our results show very good agreement with the data, suggesting that a nonvanishing gluon mass provides a better description of the spectrum of heavy mesons than the massless case. 
\end{abstract}

\maketitle

\section{Introduction}

Unraveling the infrared (IR) regime of quantum chromodynamics (QCD) is of paramount importance to grasp the mechanisms behind fundamental phenomena such as the spontaneous breaking of chiral symmetry and confinement. In contrast to the ultraviolet (UV) regime, where QCD displays asymptotic freedom, the standard Faddeev-Popov (FP) perturbative approach is ill suited for accessing the IR of the theory, due to the presence of a Landau pole. As a consequence, various nonperturbative approaches have been proposed to investigate QCD at low energies.     

Among those approaches, lattice simulations  have played a pivotal role in elucidating the intricacies of QCD in the IR regime, revealing some unexpected properties. For example, nowadays there is a consensus within the lattice community concerning the IR behavior of the gluon propagator in Landau gauge: it saturates to a finite value at vanishing momentum, similarly to a massive propagator \cite{Cucchieri:2007rg,Bornyakov:2009ug,Bogolubsky:2009dc,Maas:2011se,Oliveira:2012eh,Duarte:2016iko}. This particular behavior aligns with findings from functional methods, such as approaches rooted in Schwinger-Dyson equations (SDE) \cite{Aguilar:2008xm,Aguilar:2012rz,Huber:2018ned,Huber:2020keu} and functional renormalization group \cite{Fischer:2008uz,Cyrol:2016tym,Cyrol:2017ewj,Dupuis:2020fhh} which yield the so called decoupling solutions. Notably, preceding the confirmation through numerical simulations, such solutions had already been observed in SDE studies  \cite{Aguilar:2004sw}.

Numerical simulations have also revealed another significant property of low energy QCD: the coupling between gluons and ghosts features no singular behavior in the IR \cite{Cucchieri:2008qm,Bogolubsky:2009dc,Boucaud:2011ug}. Instead, lattice data suggest that even in the domain of very small energy, the relative order of magnitude of successive terms in perturbation theory should be of the order of 0.2 or 0.3. This finding is remarkable, implying that some sort of perturbation theory might, to some extent, capture the behavior of pure Yang-Mills theory in the IR, contradicting the results from the FP approach, which finds a Landau pole. 

This discrepancy can be attributed to the incomplete justification of the FP procedure in the IR, stemming from the presence of Gribov copies \cite{Gribov:1977wm}. Some approaches, such as the Gribov-Zwanziger (GZ) \cite{Zwanziger:1989mf} approach and its refined version \cite{Dudal:2008sp,Dudal:2010tf}, have attempted to take into account these effects by modifying the FP action and introducing various auxiliary fields.

Another possible route, instead of explaining the emergence of the gluon mass, intends to explore its consequences. One of such approaches is a particular case of a more general kind of theory put forth by G. Curci and R. Ferrari in the 1970s \cite{Curci:1976bt,Curci:1976kh} that, in Landau gauge, reduces to a gluon mass extension of the standard FP action \cite{Tissier:2010ts,Tissier:2011ey}. We will refer to this framework as the Curci-Ferrari (CF) model in Landau gauge.

The CF action is renormalizable \cite{Curci:1976bt}. In addition, as shown in \cite{Reinosa:2017qtf}, the CF model features certain trajectories of the renormalization flow that are Landau pole free, the so called infrared safe trajectories. Furthermore, as for the quenched approximation, the gauge coupling is compatible with a standard perturbative analysis for all momentum scales. By exploiting this property, several correlation functions have been evaluated at one-loop \cite{Tissier:2010ts,Tissier:2011ey,Pelaez:2013cpa,Pelaez:2014mxa,Pelaez:2015tba,Figueroa:2021sjm,Barrios:2024ixj}, displaying, in general terms, very good agreement with lattice simulations. Moreover, this agreement has improved in cases where two-loop corrections have been included \cite{Gracey:2019xom,Barrios:2020ubx,Barrios:2021cks,Barrios:2022hzr,Barrios:2024ixj}, which is consistent with the perturbative analysis.

Results of the two-point functions in the presence of quarks support the validity of the perturbative approach within the CF model for the gluon, ghost and quark dressing functions \cite{Pelaez:2014mxa,Barrios:2021cks}. As for the quark mass function, it can be reproduced with a two-loop calculation in the case of heavy quarks, whereas in the case of light quarks the model only reproduces lattice data by employing a double expansion, in the pure gauge coupling and in the inverse of the number of colors \cite{Pelaez:2017bhh,Pelaez:2020ups,Barrios:2021cks}.   

The CF model has also been employed to address phenomena of YM/QCD at finite temperature and density \cite{Maelger:2017amh,Maelger:2018vow,Maelger:2019cbk,Suenaga:2019jjv,Reinosa:2019xqq,Song:2019qoh,vanEgmond:2021jyx,Surkau:2024zfb}. In particular, it has been successfully used to investigate the QCD phase diagram \cite{Maelger:2018vow,Maelger:2019cbk,vanEgmond:2021jyx,Surkau:2024zfb} and neutron stars phenomenology \cite{Song:2019qoh}. More recently, an infrared-safe renormalization scheme has been extended to Minkowskian space-time \cite{Oribe:2025ezp}, the model has successfully been used to study confinement \cite{Pelaez:2024mtq} as well as parton distribution and fragmentation functions \cite{Bopsin:2025vhz}. For a review on the Curci-Ferrari model and its applications, see \cite{Pelaez:2021tpq}. 

Our goal in this work is to begin with the exploration of the mass spectrum of mesons by employing the CF model. To this end, heavy mesons are an excellent starting point, as they feature a rich spectroscopy, see e.g. \cite{Brambilla:2014jmp}, and can be described by nonrelativistic approaches. These include Nonrelativistic QCD (NRQCD) \cite{Caswell:1985ui,Bodwin:1994jh}, which employs an expansion in the inverse of the quark mass, and potential Nonrelativistic QCD (pNRQCD), where degrees of freedom of NRQCD that do not lead to physical states are integrated out \cite{Pineda:1997bj,Brambilla:1999xf,Brambilla:2004jw}. Heavy mesons spectra have also been obtained by means of lattice simulations, see e.g. \cite{Dudek:2007wv,Thacker:1990bm,Lepage:1992tx} and in the context of Dyson-Schwinger and Bethe-Salpeter equations (BSE), see for instance \cite{Blank:2011ha,Fischer:2014cfa,Hilger:2014nma}.

Another type of approaches, that emerged with the very discovery of charmonium, are the potential models \cite{Appelquist:1974zd,Eichten:1978tg,Godfrey:1985xj,Gupta:1993pd,Zeng:1994vj,Ebert:2002pp,Bernardini:2003ht,Cucchieri:2017icl,Mutuk:2018xay,Gutierrez-Guerrero:2021fuj}. Typically, because of the large masses of the heavy quarks, these models describe the quarkonia as a two-body system governed by the Schrödinger equation, where the quark-antiquark potential can be modeled based on phenomenological or theoretical grounds. One of the simplest models, the Coulomb plus linear potential, is simply based on the asymptotic limits of the interquark potential of QCD. More specifically, it is modeled according to a one-gluon exchange interaction at short distances and by a linear potential, responsible for confinement, at large distances \cite{Eichten:1978tg,Eichten:1979ms,Eichten:2002qv}. Several variations of this potential are available in the literature, all of them providing a good description of experimental data \cite{Eichten:1980mw,Gromes:1984ma,QuarkoniumWorkingGroup:2004kpm}. Reviews on the various theoretical and experimental approaches to heavy mesons can be found in Refs. \cite{QuarkoniumWorkingGroup:2004kpm,Brambilla:2010cs}.

In this work we aim to make a first exploration on the description of the mass spectrum of heavy mesons by the CF model. In particular, our goal is to determine whether the experimental/lattice data available for the spectra favors a nonzero gluon mass. To this end we will use a nonrelativistic potential model based on a one-massive-gluon exchange potential, which describes the short-distance regime plus a linear potential, consistent with confinement at large distances. Then, by solving the Schrödinger equation and introducing perturbative corrections, we will fit the experimental/lattice masses of bound states of charm and bottom quarks, exploring the parameter space of the model and specifically the values of the gluon mass that are compatible with the experimental/lattice data. This work complements other approaches, that also have evaluated the impact of some sort of gluon mass on heavy mesons physics, see e.g. \cite{Parisi:1980jy,Consoli:1993ew,Consoli:1997ts,Mihara:2000wf,Field:2001iu}   

It is clear that the gluon mass would exert a more substantial impact on the spectrum of relativistic mesons. However, in such cases, the spectrum must be derived by employing the BSE \cite{Salpeter:1951sz}, rendering the analysis considerably more involved. Additionally, the approximations to be used in the BSE are harder to justify. As a result, in this first exploratory paper we decided to take a simpler approach, leaving the analysis of relativistic mesons for a future work.

The article is organized as follows: In Sec. \ref{Section CF model} we introduce the Curci-Ferrari model in Landau gauge in Minkowski space-time. In Sec. \ref{Section OMGE} we present the calculation of the one-massive-gluon exchange potential, which we will incorporate into the Hamiltonian before plugging it into the Schrödinger equation. In Sec. \ref{Sec. Pert. Theory} we discuss the solutions to the Schrödinger equation as well as the introduction of perturbative corrections. In Sec. \ref{Section Renorm flow} we introduce the infrared-safe renormalization scheme of the CF model, showing the running of the gauge coupling and the gluon mass with the energy scale in the ultraviolet region. We present our results in Sec. \ref{Section Results}, discussing its relation with the gluon mass as well as the consistency with the nonrelativistic framework. Finally, in Sec. \ref{Section Discussion} we provide the conclusions and outlook of this work.


\section{The Curci-Ferrari model in Landau gauge} \label{Section CF model}
The CF model \cite{Curci:1976bt} in Landau gauge in Minkowski space-time is defined by the following Lagrangian density,

\begin{align}
  \label{eq_lagrang}
  \mathcal L&=-\frac 14 F_{\mu\nu}^a F^{\mu\nu,a}-h^a\partial^\mu A_\mu^a+\frac {m^2}2 A_\mu^a A^{\mu,a}\\
  &-\partial ^\mu\overline c^a(D_\mu
  c)^a+\sum_{j=1}^{N_f} \bar{\psi_j}(i\slashed{D}-M_j)\psi_j \nonumber,
\end{align}
where, as usual, we introduced $(D_\mu c)^a=\partial_\mu c^a+ g f^{abc}A_\mu^b c^c$ and the field strength tensor $F_{\mu\nu}^a=\partial_\mu A_\nu^a-\partial_\nu
A_\mu^a+gf^{abc}A_\mu^bA_\nu^c$. The parameter $g$ refers to the bare gauge coupling and Latin indices are associated with color indices of the SU(3) gauge group. The mass parameters $m$ and $M_j$ are the bare gluon and quark mass for each flavor $j$, respectively. 

In the context of the investigation of the IR regime of QCD, the CF model is strongly motivated by the decoupling behavior of the gluon propagator in lattice simulations \cite{Bowman:2007du,Bogolubsky:2009dc,Oliveira:2012eh,Duarte:2016iko,Oliveira:2018lln}. In this respect, the CF model is the simplest renormalizable deformation of the standard FP action which, by explicitly adding a gluon mass term, is able to account for these results \cite{Tissier:2010ts,Tissier:2011ey}. Being the gluon field massive, the tree level gluon propagator in momentum space reads 

\begin{equation}
  G_{\mu \nu}^{a b}(q)=\frac{-i\delta^{ab}}{q^2-m^2+i\epsilon} \left( \eta_{\mu \nu}-\frac{q_\mu q_\nu}{q^2}\right).  \label{gluonpropagator}
\end{equation}

Before ending this section is important to stress that the gluon mass term is introduced at the level of the gauge-fixed action and, up to date, has not been derived from first principles. We believe, however, that at least partially this term should capture the effect of the Gribov copies in the IR\footnote{See, for instance, Ref. \cite{Serreau:2012cg} where the gauge fixing procedure introduced by the authors, which averages over the Gribov copies, lifts the Gribov degeneracy and leads to a massive extension of the FP action. However, the relation between this term and the CF gluon mass is not clear.}. Whatever the gluon mass origin may be, we will consider the CF model as a phenomenological tool, which has shown to provide very good results while introducing a minimal amount of modeling. 


\section{One-massive gluon exchange potential} \label{Section OMGE}

The Bethe-Salpeter equation (BSE) \cite{Salpeter:1951sz} provides a covariant relativistic formalism to describe two-particle bound states by including a kernel that encompasses all two-particle-irreducible interactions between the constituents. A rigorous (semi)analytic description of the quarkonium spectrum typically requires solving the BSE. However, in the nonrelativistic limit, as the works in the context of potential models have demonstrated \blue{\cite{Appelquist:1974zd,Eichten:1978tg,Godfrey:1985xj,Gupta:1993pd,Zeng:1994vj,Ebert:2002pp,Bernardini:2003ht,Cucchieri:2017icl,Mutuk:2018xay,Gutierrez-Guerrero:2021fuj}}, heavy quarkonium is well described by the simpler Schrödinger equation. 

To evaluate the potential, given the exploratory nature of this work, we will restrict ourselves to next-to-leading order corrections within the one-gluon exchange approximation. A rigorous analysis regarding all next-to-leading order diagrams in the CF framework goes beyond the scope of the present work\footnote{Such analysis should include a detailed study of the order of each diagram contributing to the meson masses, putting emphasis on the enhancement of the diagrams due to the presence of complex poles.}.  


\begin{figure}[h]
\begin{centering}
\includegraphics[scale=0.90]{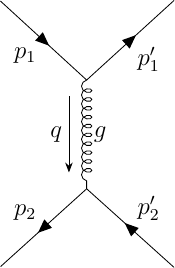}
\par\end{centering}
\caption{Tree-level diagram in QCD for quark-antiquark state transition}
\label{Diagram}
\end{figure}

Thus, we derive the quark-antiquark interaction potential from the scattering amplitude of a one-gluon exchange diagram from the CF model. Following Born’s approximation \cite{Lucha:1995zv} this potential is, up to a prefactor, given by the Fourier transform of the scattering amplitude depicted in Fig. \ref{Diagram}. The diagram represents a tree-level process in which a quark with initial momentum $p_1$ interacts with an antiquark of momentum $p_2$  through the exchange of a gluon carrying momentum $q$. Primed quantities indicate the final momentum of the particles involved. Such a potential will be composed of a leading Yukawa term and relativistic next-to-leading order corrections, including spin-dependent interactions, as we will see.\\



Applying the Feynman rules to the tree-level diagram shown in \cref{Diagram}, we obtain the scattering matrix (S-matrix) of the process:
\begin{equation}
\begin{aligned}
i \mathcal{M} = [ \bar{u}(p_1'){{c_1'}}^\dagger & (ig {\gamma}^{\mu} t^a)c_1 u(p_1)] G_{\mu\nu}^{ab}(q) \\ & 
[ \bar{v} (p_2)c_2^\dagger (ig \gamma^\nu t^b)c_2'v(p_2') ].
    \label{amp}
\end{aligned}
\end{equation}

\noindent Here, 
\begin{align}
u(p_1) = &\;\sqrt{2m}\begin{pmatrix} \left(1 - \frac{\mathbf p_1^2}{8M_1^2}\right)w_1 \\
			\frac{\mathbf \sigma_1 \cdot \mathbf p_1}{2M_1}w_1
		\end{pmatrix},  \\
\bar{u}(p'_1) = & \sqrt{2M_1}\begin{pmatrix}
			{w'}_1^{\dagger} \left(1 - \frac{\mathbf {p'}_1^2}{8M_1^2}\right) & -{w'}_1^{\dagger}
			\frac{\mathbf{\sigma_1} \cdot \mathbf {p'_1}}{2M_1}
		\end{pmatrix},
		\label{espinores_fermion}
\end{align}
are the Dirac spinors associated with quarks, and 
\begin{equation}
\begin{aligned}
		v(p'_2) = &\;\sqrt{2M_2}\begin{pmatrix}
			\frac{\boldsymbol \sigma_2 \cdot \mathbf p'_2}{2M_2}w'_2 \\
			\left(1 - \frac{\mathbf {p'}_2^2}{8M_2^2}\right){w'}_2
		\end{pmatrix},  \\
		\bar{v}(p_2) = & \sqrt{2M_2}	\begin{pmatrix}
	w_2^\dagger\frac{\boldsymbol \sigma_2 \cdot \mathbf p_2}{2M_2}
	& -w_2^\dagger\left(1 - \frac{\mathbf p_2^2}{8M_2^2}\right)
	\end{pmatrix},
\end{aligned}
\end{equation}
are those associated with antiquarks.
Due to momentum conservation the initial and final quantities are related by $\mathbf{p}_1'=\mathbf{p}_1-\mathbf{q}$ and $\mathbf{p}_2'=\mathbf{p}_2+\mathbf{q}$.


There are several ways to determine the component $q_0$ of the gluon momentum. One of them is to consider that the external particles are on shell. This choice is specially useful when dealing with scattering amplitudes. In contrast, as for bound states a more natural choice is $q_0=0$. This is because the location of bound states poles in the truncated two-particle Green's function is independent of the relative energies of the constituents \cite{Caswell:1978mt}\footnote{Furthermore, in \cite{Caswell:1978mt} it is shown that, provided that the kernel of the BSE is static, the truncated two-particle Green's function is actually independent of $q_0$.}.

Finally, the amplitude \eqref{amp} takes the form

\begin{equation}
    \begin{aligned}
	i\mathcal M = g^2  \frac{4}{3} &\frac{i 4 M_1 M_2}{(\mathbf q^2 + m^2)}    
	\left \lbrace -1 + \left( \frac{1}{M_1^2} + \frac{1}{M_2^2} \right)  \frac{\mathbf q^2}{8} \right.  \\ &
 +\frac{i\boldsymbol \sigma_1 \cdot (\mathbf q \times \mathbf p_1)}{4M_1^2} + \frac{i\boldsymbol \sigma_2 \cdot (\mathbf q \times \mathbf	p_2)}{4M_2^2} \\ &
	+ \frac{\mathbf p_1 \cdot \mathbf p_2}{M_1M_2} - \frac{(\mathbf p_1 \cdot \mathbf q)(\mathbf p_2
	\cdot \mathbf q)}{M_1M_2\mathbf q^2} \\ &
 - \frac{i\boldsymbol \sigma_1 \cdot (\mathbf q 
	\times \mathbf p_2)}{2M_1M_2} -
	\frac{i\boldsymbol \sigma_2 \cdot (\mathbf q \times \mathbf
	p_1)}{2M_1M_2} \\ &
        \left.  - \frac{(\boldsymbol \sigma_1
	\cdot \boldsymbol \sigma_2)\mathbf q^2}{4M_1M_2} +  \frac{(\boldsymbol \sigma_1
	\cdot \mathbf q )(\boldsymbol \sigma_2\cdot \mathbf q)}
	{4M_1M_2} \right \rbrace
	\label{amplitud_charmbott}
\end{aligned}
\end{equation}

As already mentioned, the interaction potential is found by taking the Fourier transform of the amplitude \eqref{amplitud_charmbott} (without the factor $i 4 M_1 M_2$). The details of this calculation are shown in \cref{appendix_A_Fourier}. As a result, we obtain the potential

\begin{equation}
H= \sum_{i=0}^{i=9} H_i
\label{hamiltonian}
\end{equation}
where
\begin{equation}
\begin{aligned}
    H_0 &= \frac{1}{2\mu}p^2 -g^2\frac{e^{-mr}}{3\pi r} \label{Ho}
\end{aligned}
\end{equation}
and $\mu\equiv\frac{M_1M_2}{M_1+M_2}$ is the reduced mass of the meson. Besides,
\begin{equation}
        H_1 = - \left( \frac{1}{M_1^3} + \frac{1}{M_2^3} \right) \frac{\mathbf{p}^4}{8},
        \label{eq:H1}
\end{equation}

\begin{align}
H_2= - \left( \frac{1}{M_1^2} + \frac{1}{M_2^2} \right)  \frac{g^2 m^2 e^{-mr}}{24\pi r},
\end{align}

\begin{align}
H_3= \frac{g^2}{6} \left( \frac{1}{M_1^2} + \frac{1}{M_2^2} \right)\delta^3(r),
\end{align}

\begin{align}
     H_4&=g^2 (mr+1) \frac{e^{-mr}}{3\pi r^3} \left( \frac{1}{2M_1^2} + \frac{1}{M_2M_1} \right) \boldsymbol{s_1} \cdot \mathbf L  \nonumber\\
     &
     +g^2 (mr+1) \frac{e^{-mr}}{3\pi r^3} \left( \frac{1}{2M_2^2} + \frac{1}{M_2M_1} \right) \boldsymbol{s_2} \cdot \mathbf L , 
\end{align}

\begin{equation}
    \begin{aligned}
        H_5 =  \frac{g^2 }{3\pi M_1M_2m^2} \biggl[ \frac{\mathbf {p}^2}{r^3} -  \frac{3 (\mathbf p \cdot \mathbf{r} )^2}{r^5} \biggr ],
    \end{aligned}
\end{equation}

\begin{equation}
    \begin{aligned}
        H_6 =&  \frac{g^2 e^{-mr} }{3\pi M_1M_2m^2} \left\lbrace - \frac{\mathbf {p}^2}{r^3} (mr+1)  \right. \\ &
 + \frac{3(\mathbf p \cdot \mathbf{r} )^2}{ r^5} \left( \frac{m^2r^2}{3}+mr+1 \right) \biggr \rbrace,
    \end{aligned}
\end{equation}

\begin{equation}
\begin{aligned}
     H_7= \frac{8}{9} \frac{g^2}{M_1M_2} (\boldsymbol s_1 \cdot \boldsymbol s_2) \delta^3(r),
\end{aligned}
\end{equation}

\begin{equation}
\begin{aligned}
     H_8= -\frac{g^2e^{-mr}}{3\pi M_1M_2} \frac{\boldsymbol s_1 \cdot \boldsymbol s_2}{r^3}(m^2r^2+mr+1) ,
\end{aligned}
\end{equation}

\begin{equation}
\begin{aligned}
     H_9=& \frac{g^2e^{-mr}}{3\pi M_1M_2} \frac{(\boldsymbol s_1 \cdot \mathbf{r}) (\boldsymbol s_2 \cdot \mathbf{r})}{r^5} \left( \frac{m^2r^2}{3}+mr+1\right)
     \label{eq:H9}
\end{aligned}
\end{equation}
\\
As we will work in the center of mass frame of the system, the momenta of the particles satisfy  $\mathbf p = \mathbf p_1 = - \mathbf p_2$. The orbital angular momentum is $\mathbf L = \mathbf r \times \mathbf p$, and the spin operators of the quark and antiquark (represented by $\mathbf{s}_{1}$ and $\mathbf{s}_{2}$, respectively) are related to the Pauli matrices as follows 
\begin{equation}
\boldsymbol{s}_{1}=\frac{\boldsymbol{\sigma}_{1}}{2}, \hspace{0.5cm} \boldsymbol{s}_{2}=-\frac{\boldsymbol{\sigma}_{2}}{2}. \label{sigma-spin}
\end{equation}

The Hamiltonian \eqref{hamiltonian}, associated with the two-body problem, agrees with that obtained by Gromes in \cite{Gromes:1976np} for a two-fermion system of equal masses.
We note that the spin terms are of Breit-type, as can be seen in Refs. \cite{Breit1, Breit2, Breit3} as well as in an example of Breit interaction, provided in Ref. \cite{Eiglsperger:2007ay}.

\section{Solutions to the Schrödinger equation and perturbative corrections} \label{Sec. Pert. Theory}

The Hamiltonian \eqref{hamiltonian} is nonconfining. To take into account confinement, we can add to this Hamiltonian a linear potential \cite{Eichten:1979ms,Eichten:2007qx} of the type 
\begin{equation}
    V_b=b r.
\end{equation}
This is the Cornell potential, where the parameter $b$ is called the string tension. Having our confining potential, we can proceed now to determine the mass spectrum of quarkonium. To this end we solve numerically the Schrödinger equation, with a Hamiltonian given by 
\begin{align}
    H_S = \frac{1}{2\mu}\bold{p}^2 -g^2\frac{e^{-mr}}{3\pi r} + b r, \label{Ho+Vb}
\end{align}
where the first two terms correspond to the leading contributions to the Hamiltonian \eqref{hamiltonian}, whereas the third term is the nonperturbative Cornell potential. The numerical solutions of the equation will give us eigenstates and eigenenergies of $H_S$. Afterward, fine corrections are introduced perturbatively. 

As we are working with a central potential, the states in the position representation read
\begin{equation}
	\psi_{n\ell m}(r) = R_{n\ell}(r) Y_\ell^m(\theta,\phi),
\end{equation}
where $Y_\ell^m(\theta,\phi)$ are spherical harmonics with eigenvalues $\ell(\ell+1)$ of $\mathbf{L}^2$ (where $\ell= 0, 1/2, 1, 3/2, 2...$) and eigenvalues $m$ of $L_z$. Now, we can write the radial part of the Schrödinger equation (in analogy to the case of the hydrogen atom) as

\begin{align}
\frac{-1}{2\mu r^2} \frac{d}{dr} \left( r^2 \frac{dR_{n\ell}}{dr} \right)+\biggl[\frac{\ell(\ell+1)}{2\mu r^2}+&V(r)\biggr]R_{n\ell}
=E_{n\ell} R_{n\ell}
 \label{schro_r}
\end{align}
where $\ell$ is the angular momentum quantum number and  $n$ is the index associated with the energy level. \\

In order to express \cref{schro_r} in terms of dimensionless quantities for distance and energy, we redefine the radial variable as $\rho=\frac r{a_0}$, with $a_0=\frac{3\pi}{g^2\mu}$, and the energy eigenvalues as ${\lambda_{n\ell}=\sqrt{-\frac{E_{n\ell}}{E_I}}}$, with ${E_I  = \frac{g^4\mu}{2(3\pi)^2}}$. 
As a result, \cref{schro_r} transforms into:

\begin{equation}
	 \left[\frac{d^2 }{d\rho^2} - 
	  \frac{\ell(\ell+1)}{\rho^2}+\frac{2 e^{-\tilde{m} \rho}}
	 {\rho}+\tilde{b}\rho -\lambda^2_{n\ell} \right]u_{n\ell}(\rho)=0,
    \label{schro_u}
\end{equation}
with $\tilde{b} = \frac{b a_0}{E_I}$ and $\tilde{m}=ma_0$. The function $u_{n\ell}$ is defined in terms of $R_{n\ell}(r)$ as $u_{n\ell}(r)=r R_{n\ell}(r)$, and the condition at the origin is $u_{n\ell}(0)=0$.
\cref{schro_u} can be numerically solved to find the spectrum and the unperturbed eigenstates. The corrections to the energy can be calculated as
\begin{equation}
    E_1= \langle n_0|H_P|n_0 \rangle,
    \label{eq:pert_corr}
\end{equation}
where $\left|n_0\right \rangle$ are the eigenstates of $H_S$ and $H_P=\sum_{i=1}^{i=9}H_i$, where the addends $H_i$ have been defined in Eqs. \eqref{eq:H1} to \eqref{eq:H9}. 
We will work with the states $n=1$, $\ell=0$, $\ell=1$. With the purpose of computing the correction $E_1$, the Hamiltonian $H_P$ must be diagonalizable in the basis $\{|n_0\rangle\}$. The details regarding the choice of the appropriate basis for diagonalizing the perturbative terms can be found in \cite{Lucha:1991vn,Cahn:2003cw}. The evaluation of $E_1$ is performed numerically, after calculating the Fourier transform of $H_P$, as explained in App. \ref{app corrections}. 

\section{Renormalization group} \label{Section Renorm flow}
Among other quantities, our results will depend on the gauge coupling $g$ and the gluon mass $m$. Since the typical scales associated with charmonium, charm-bottom mesons and bottomonium differ from each other by various GeVs, it is important to introduce the dependence on the scale of these parameters. Within the CF model, this dependence has already been derived from the evaluation of the two-point functions of the model, both at one- and two-loop order in pure YM theory \cite{Tissier:2011ey,Gracey:2019xom} and QCD \cite{Pelaez:2014mxa,Barrios:2021cks}. To write down the renormalization conditions implemented in the latter case, it is convenient to introduce the renormalization, or $Z$-, factors for the fields,
\begin{align}
    A^{a,\mu}_B=\sqrt{Z_A}A^{a,\mu}&,\quad c^a_B=\sqrt{Z_c} c^a,\quad \bar{c}^a_B=\sqrt{Z_c}\bar{c}^a,\nonumber \\ \psi_B&=\sqrt{Z_\psi}\psi,\quad \bar{\psi}_B= \sqrt{Z_\psi}\bar{\psi},
   \end{align}
and the parameters
\begin{equation}
    g_B=Z_g g \ \quad, m^2_B=Z_{m^2}m^2,\quad M_B=Z_M M.
\end{equation}
In the above notation, bare and renormalized quantities differ by the label `B', which denotes the former. In the end, there are six independent renormalization factors that are determined through the following renormalization conditions:
\begin{align}
    &G^{-1}(p=\tilde{\mu})=m^2+\tilde{\mu}^2,\quad F(p=\tilde{\mu})=1,\nonumber \\
    &Z(p=\tilde{\mu})=1, \quad M(p=\tilde{\mu})=M.
    \label{eq:irs_2pointfunctions}
\end{align}
and
\begin{equation}
    Z_g \sqrt{Z_A}Z_c=1,\quad Z_{m^2}Z_A Z_c=1.
    \label{eq:irs_theo}
\end{equation}
The conditions provided in \eqref{eq:irs_2pointfunctions} impose that the gluon propagator, $G$, the ghost dressing function, $F$, and the form factors associated with the quark propagator
\begin{equation*}
    S(p)=Z(p)\frac{i \slashed p +M(p)}{p^2+M^2(p)}
\end{equation*}
acquire their tree-level form at the renormalization scale $p=\tilde{\mu}$. In the case of the function $M(p)$, we impose it to equal the quark mass at the renormalization scale. The conditions from \eqref{eq:irs_theo} extend two nonrenormalization theorems of the CF model to the finite parts of the renormalization factors. The set of six conditions outlined above defines an infrared safe renormalization scheme, meaning that the coupling does not feature a Landau pole in the infrared. In order to find the running of $g$ and $m$ let us introduce their respective anomalous dimensions:
\begin{equation}
    \gamma_{m^2}\equiv\frac{d \ln Z_{m^2}}{d\ln \tilde{\mu}},\quad 
    \gamma_{g}\equiv\frac{d \ln Z_{g}}{d\ln \tilde{\mu}}
\end{equation}
and the $\beta-$functions,
\begin{equation}
    \beta_{m^2}\equiv\frac{d m^2}{d\ln \tilde{\mu}},\quad 
    \beta_g\equiv \frac{d g}{d\ln \tilde{\mu}}.
\end{equation}
The dependence of the parameters on the scale is encoded in the $\beta-$functions, that relate to the anomalous dimensions as \cite{Barrios:2021cks}
\begin{equation}
    0=\gamma_{m^2}+\frac{\beta_{m^2}}{m^2}=\gamma_g+\frac{\beta_g}{g}.
\end{equation}
As a result, the $\beta$-functions can be extracted directly from the $Z$-factors. As we are interested in scales that range in between 3 and 10 GeV, we will use the leading ultraviolet contribution to the one-loop results. In this approximation, the $\beta$-functions take the form \cite{Pelaez:2014mxa,Barrios:2021cks}
\begin{align}
\beta_{m^2}^{\text{UV}}&=\frac{m^2 g^2}{16\pi^2}\left(-\frac{35}{6}N_c+\frac 4 3 N_f \right),\label{eq:betaUV1}\\
\beta_g^{\text{UV}}&=-g^3 \beta_0;\hspace{1mm}\beta_0=\frac{1}{16\pi^2}\left(\frac{11}{3} N_c-\frac{2}{3}N_f \right).
\label{eq:betaUV2}
\end{align}
where $N_c=3$ is the number of colors and $N_f=4$ is the number of active quark flavors at scales relevant for mesons composed of charm and bottom quarks. From \cref{eq:betaUV1,eq:betaUV2} we can finally obtain the evolution of the parameters with the energy scale $\tilde{\mu}$:
\begin{equation}
    g^2(\tilde{\mu})=\frac{g_0^2}{1+g_0^2 \beta_0\ln (\tilde{\mu}^2/\tilde{\mu}_0^2)},
    \label{eq:lambda_run}
\end{equation}
and
\begin{equation}
    m^2(\tilde{\mu})=m_0^2\left(\frac{g(\tilde{\mu})}{g_0}\right)^\kappa,
    \label{eq:mass_run}
\end{equation}
where $\kappa=\frac{35N_c-8N_f}{2(11N_c-2N_f)}$, $m_0\equiv m(\tilde{\mu_0})$, $g_0=g(\tilde{\mu_0})$ and $\tilde{\mu_0}$ is some initialization scale. In this work, we focus on three distinct energy scales, each associated with charmonium, charm-bottom mesons and bottomonium. In principle, we could consider a similar dependence on the energy scale for the quark masses. However, the quantity actually relevant for our calculation is the quark pole mass, which does not feature such dependence.

\section{Results} \label{Section Results}

Our goal is to fit the experimental/lattice data of the spectrum of quarkonia, composed exclusively of charm and bottom quarks and their respective antiquarks, identifying the sets of parameters compatible with the data\footnote{The experimental data is extracted from \cite{ParticleDataGroup:2024cfk}. In the case of the meson $B_c^*$, as there is no experimental data available for its mass spectrum, we compared it with the result from lattice simulations \cite{Mathur:2018epb}. Our results do not change significantly when excluding this case from the fit.}. Specifically, we aim to determine whether the case of nonvanishing gluon mass is preferred over the massless case when fitting the experimental/lattice data. We restrict the analysis to ground states, $n=1$, with angular momentum $\ell=0$ and $\ell=1$, because it is in the case of these quantum numbers that we expect the nonrelativistic approximation to work best.

To fit the data, first we concentrate on the bottomonium spectrum, since the nonrelativistic approximation is expected to be more reliable in this case than for charmonium. We investigate the sets of parameters compatible with the experimental data, determining the value of the bottom mass. Then, we determine the remaining parameters, by fitting the spectrum of all the heavy mesons together. 

We compute the bottomonium spectrum for a range of values of the parameters of the model, $\{b,M_b,g(\mu_b),m(\mu_b)\}$, where $b$ is the string tension, $M_b$ the bottom mass, $g(\mu_b)$ and $m(\mu_b)$ are the coupling and the gluon mass, evaluated at the scale of the reduced mass of the two-body system made of a bottom and an antibottom quark, i.e. $\mu_b=\frac{M_b}{2}$. Once the spectrum is known for a given set of parameters, we calculate the error with respect to the experimental results: 
\begin{equation}
\chi_b=  \sqrt{ \frac{1}{N} \sum^N_{i=1} \biggl( \frac{M^{\text{CF}}_i-M^{\text{exp}}_i}{M^{\text{exp}}_i} \biggr )^2},
\end{equation}
where the index $i$ runs over the number of heavy mesons of the type $\bar{b}b$ and $M_i^{\text{CF}}$ and $M_i^{\text{exp}}$ are the theoretical and experimental values of a meson mass, respectively. We implemented the following parameter scans: for the string tension, $b$, a scan from $0.02$ GeV$^{2}$ to $0.5$ GeV$^{2}$ with a step size of $db=0.01$ GeV$^{2}$; for the bottom quark mass, $M_b$, a scan from $4.0$ GeV to $6.5$ GeV, with a step size of $dM_b=0.05$ GeV; for the coupling, $g(\mu_b)$, a scan from $1.4$ to $1.9$, with a step size of $dg=0.05$; for the gluon mass, $m(\mu_b)$, a scan from $0.02$ GeV to $0.5$ GeV, with a step size of $dm=0.02$ GeV. Since the value of the bottom mass that best fits the data lay between $M_b=4.55$ GeV and $M_b=5.00$ GeV, we also did the evaluation for $M_b=4.575$ GeV. The values of the parameters that best fit the data for the bottomonium spectrum are $\{b,M_b,g(\mu_b),m(\mu_b)\}=\{0.33 ~ \text{GeV}^2,4.575~ \text{GeV},1.75,40\ \text{MeV}\}$, corresponding to an error of $\chi_b=0.0004895$. 

Then, we proceed as follows. First, we extract the value of $M_b$ directly from the fit of the bottomonium spectrum, i.e. $M_b=4.575$ GeV. The remaining parameters $\{b,M_c,g(\mu_c),m(\mu_c)\}$ are determined by fitting the spectra of bottomonium, charmonium and charm-bottom mesons together\footnote{Of course, $M_c$ depends only on the spectrum of the charmonium and charm-bottom mesons.}. When fitting all the heavy mesons, the gluon mass and the coupling are evaluated at the scale $\mu_c=\frac{M_c}{2}$\footnote{It is clear from \cref{eq:lambda_run,eq:mass_run} that the running of the parameters depend only on the quotient $\tilde{\mu}/\tilde{\mu}_0$. This means that we would obtain the same results for scales $\mu_b=\alpha M_b$ and $\mu_c=\alpha M_c$, with $\alpha>0$.}. We performed the following scans: for the string tension, $b$, the same scan as the one implemented for the bottomonium spectrum; for the charm mass, $M_c$, a scan from $1.1$ GeV to $1.3$ GeV, with a step size of $dM_c=0.05$ GeV; for the coupling, $g(\mu_c)$, a scan from $1.4$ to $2.3$, with a step size of $dg=0.05$; for the gluon mass, $m(\mu_c)$, a scan from $0.02$ GeV to $0.8$ GeV, with a step size of $dm=0.01$ GeV. 

To fit the data we minimized the error $\chi$, whose definition is analogous to that of $\chi_b$, but extended to all the heavy mesons studied in this paper. The values that fit the experimental/lattice data are $\{b,M_c,M_b,g(\mu_c=600\text{MeV}),m(\mu_c=600\text{MeV})\}=\{0.26\ \text{GeV}^2,1.20\ \text{GeV},4.575\ \text{GeV}, 2.05,370\ \text{MeV}\}$. At scales relevant for bottomonium, by employing \cref{eq:lambda_run,eq:mass_run}, the values of the gluon mass and the coupling are $m(\mu_b)=312$ MeV and $g(\mu_b)=1.62$, respectively. The values of $M_c$ and $M_b$ are consistent with those reported in the literature, see, for instance, \cite{QuarkoniumWorkingGroup:2004kpm} and the references therein. As for the string tension, $b$, it is somewhat higher than usually reported \cite{QuarkoniumWorkingGroup:2004kpm,Deur:2016tte} but still compatible with some studies, see, for example \cite{Ding:1995he}. In Refs. \cite{Tissier:2011ey,Pelaez:2014mxa,Gracey:2019xom,Barrios:2021cks}, the gluon mass and the coupling from the CF model were estimated by fitting two-point correlation functions from lattice simulations. The values presented in those works correspond to the scale $\mu_0=$ 1 GeV. In this work we get, $g(\mu_0=1\ \text{GeV})=1.75$ and $m(\mu_0=1\ \text{GeV})=330$ MeV, which are compatible with the results of those references.

\begin{figure}[h]
\centering
\begin{subfigure}[t]{0.5\textwidth}
\hspace{-0.2cm}\includegraphics[scale=0.7]{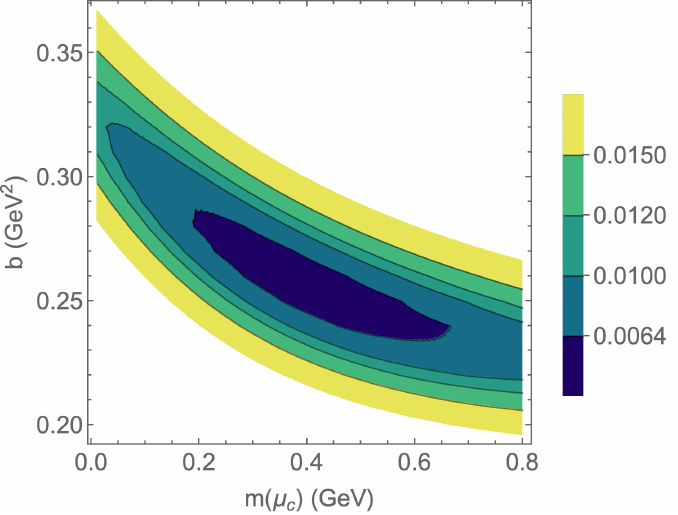}
\end{subfigure}
\begin{subfigure}{0.5\textwidth}
\includegraphics[scale=0.7]{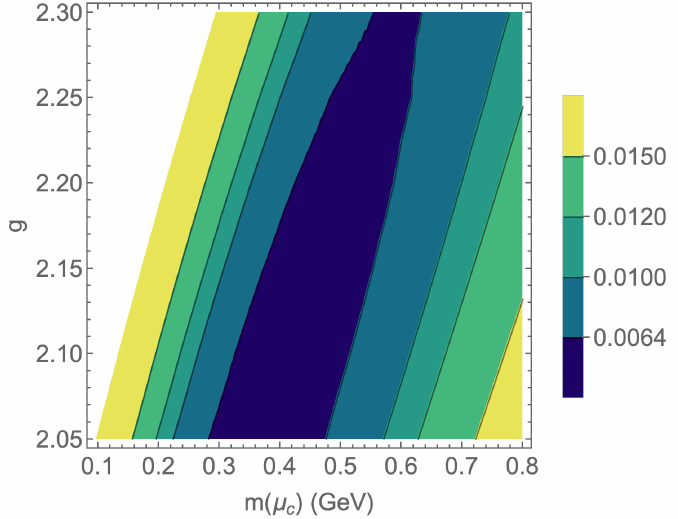}
\end{subfigure}
\caption{Contour lines of the error $\chi$ corresponding to $\{g(\mu_c),M_c,M_b\}=\{2.05,1.20~ \text{GeV}, 4.575~ \text{GeV}\}$ (top) and $\{b,M_c,M_b\}=\{0.26~ \text{GeV}^2, 1.20~ \text{GeV}, 4.575~ \text{GeV}\}$ (bottom).}
\label{fig:contour_plots_comp_error}
\end{figure}

Contour plots of the error $\chi$ as a function of the parameters are displayed in \cref{fig:contour_plots_comp_error}. We observe that the region of the parameter space with minimum $\chi$ corresponds to a nonzero gluon mass. However, in order to know whether our results actually rule out the massless case, we need to estimate the uncertainty of our computation. To that end, we note that the meson masses from the CF model, $\mathrm{M}_{\text{th.}}$, read
\begin{equation}
    \mathrm{M}_{\text{th.}}=\mathrm{M}_{0}+\lambda \mathrm{M}_{1}+\lambda^2 \mathrm{M}_{2}+\mathcal{O}(\lambda^3), 
\end{equation}
where $\lambda\equiv \frac{g^2 N_c}{16\pi^2}$ is the small parameter of the perturbative expansion in the CF model. Thus, $\chi^2$ can be estimated as
\begin{equation}
    \chi^2\sim\frac{1}{N} \sum^N_{i=1} \lambda^4  \left(\frac{\mathrm{M}_{2}}{\mathrm{M}_{i,\text{exp}}}\right)^2\sim \lambda^4,  
\end{equation}
where in the last step we assumed that the terms $\mathrm{M}_{2}$ and $\mathrm{M}_{i,\text{exp}}$ are of the same order. This means that all those sets of parameters $\{b,M_c,M_b,g(\mu_c),m(\mu_c)\}$ whose error is such that $\chi\lesssim \lambda^2=0.006374$ are compatible with the experimental/lattice results. It is clear from the plots of \cref{fig:contour_plots_comp_error} that a vanishing gluon mass is safely outside of this region. In principle, we could conclude that the massless gluon case is not compatible with our results. Nevertheless, as we observed that relatively small changes in the values of the parameters can lead to modifications on the contour plots, we need further studies, such as the analysis of other observables, before reaching a robust conclusion on this matter. We plan to do this investigation in a future work.

\begin{figure*}[t]
\centering
\begin{subfigure}[b]{\textwidth}
    \includegraphics[scale=0.8]{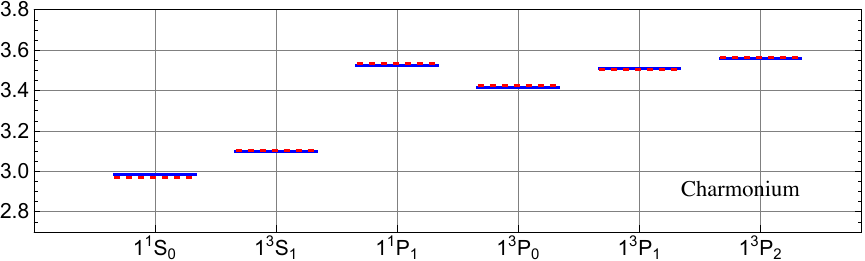}
\end{subfigure}
\hspace{0.4cm}
\begin{subfigure}[b]{\textwidth}
er    \includegraphics[scale=0.8]{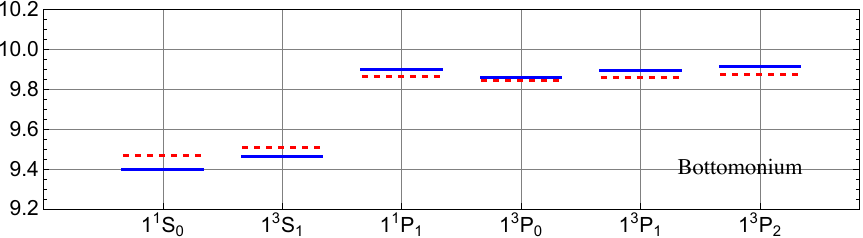}
\end{subfigure}
\begin{subfigure}[b]{\textwidth}
    \includegraphics[scale=0.7]{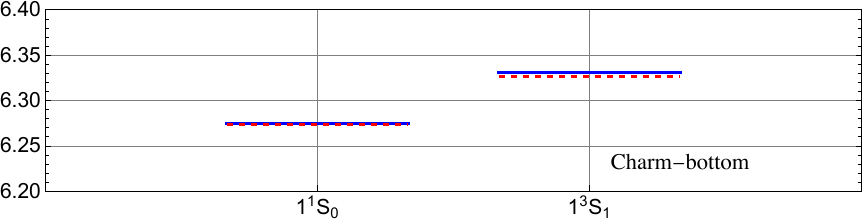}
\end{subfigure}
\hspace{0.4cm}
\caption{Comparison between the experimental/lattice data and the CF results for the spectrum of charmonium (top), bottomonium (middle) and the $B_c$ meson spectrum (bottom). The vertical axis refers to mass values in GeV. In all cases blue solid lines correspond to experimental data (lattice data in the case of $B_c^*$) and red lines to our results.}
\label{fig:comparison_experiments}
\end{figure*}

The results of our computation and the experimental/lattice data of the meson masses are provided in \cref{tab:ccbar,tab:bc,tab:bbbar}. Graphically, our results for the spectra are displayed in \cref{fig:comparison_experiments}. In all cases, our results are in good agreement with experimental/lattice data. As for charmonium and charm-bottom mesons, our results and the experimental/lattice data agree within the margins of error. This is not the case for the bottomonium spectrum, where it is clear that we are underestimating the error of our procedure. 

\begin{table}[t]
    \centering
    \begin{tabular}{|| c|c|c|c ||}
    \hline \hline
      State &  \makecell{Particle\\ name} & \makecell{Experimental\\ mass (GeV)} & \makecell{Calculated mass \\ (this work) (GeV)}\\
      \hline \hline
         $1^1S_0$   & $\eta_c$  & 2.9841(4) & 2.9684(189) \\
         $1^3S_1$ & J/$\psi$ & 3.096900(6) & 3.1014(198) \\
         \hline
         $1^1P_1$ & $h_c$ & 3.52537(14) & 3.5334(225) \\
         $1^3P_0$ & $\chi_{c0}$ & 3.41471(30) & 3.4262(218) \\
         $1^3P_1$ & $\chi_{c1}$ & 3.51067(5) & 3.5051(223) \\
         $1^3P_2$ & $\chi_{c2}$ & 3.55617(7) & 3.5645(227)\\
         \hline \hline
    \end{tabular}
    \caption{Experimental \cite{ParticleDataGroup:2024cfk} and calculated masses (this work) for $c \bar{c}$ states.}
    \label{tab:ccbar}
\end{table}

\begin{table}[h]
    \centering
    \begin{tabular}{|| c|c|c|c ||}
    \hline \hline
      State &  \makecell{Particle\\ name} & \makecell{Experimental - Lattice\\ mass (GeV)} & \makecell{Calculated mass \\ (this work) (GeV)}\\
      \hline \hline
         $1^1S_0$   & $B_c$  & 6.27447(32) & 6.2730(258) \\
         $1^3S_1$ & $B_c^*$ & 6.331(4)(6) & 6.3263(278) \\
         \hline \hline
    \end{tabular}
    \caption{Experimental \cite{ParticleDataGroup:2024cfk} mass of $B_c$, lattice result \cite{Mathur:2018epb} of the mass $B_c^*$ and calculated masses (this work). }
    \label{tab:bc}
\end{table}

\begin{table}[t]
    \centering
    \begin{tabular}{|| c|c|c|c ||}
    \hline \hline
      State &  \makecell{Particle\\ name} & \makecell{Experimental\\ mass (GeV)} & \makecell{Calculated mass \\ (this work) (GeV)}\\
      \hline \hline
         $1^1S_0$   & $\eta_b$  & 9.3987(20) & 9.4692(238) \\
         $1^3S_1$ & $\Upsilon$ & 9.46040(10) & 9.5063(239) \\
         \hline
         $1^1P_1$ & $h_b$ & 9.8993(8) & 9.8647(248) \\
         $1^3P_0$ & $\chi_{b0}$ & 9.85944(42)(31) & 9.8425(247) \\
         $1^3P_1$ & $\chi_{b1}$ & 9.89278(26)(31) & 9.8589(247) \\
         $1^3P_2$ & $\chi_{b2}$ & 9.91221(26)(31) & 9.8719(248)\\
         \hline \hline
    \end{tabular}
    \caption{Experimental \cite{ParticleDataGroup:2024cfk} and calculated masses (this work) for $b \bar{b}$ states.}
    \label{tab:bbbar}
\end{table}

Before ending this section, it is important to crosscheck the self-consistency of our approach by providing an estimate of the velocities of the valence quarks of the mesons. As explained in \cite{Domenech-Garret:2008itl,Debastiani:2017msn}, this can be calculated as

\begin{equation}
    \langle v^2 \rangle=\frac{1}{2\mu}(E-\langle V^{(0)}(r)\rangle). 
\end{equation}
In this evaluation, we restrict our calculation to the potential $V^{(0)}(r)=-g^2\frac{e^{-mr}}{3\pi r}+b r$, where $\mu$ is the reduced mass of the system. Of course, in this estimate we are not taking into account the corrections from Eqs. \eqref{eq:H1} to \eqref{eq:H9}. Thus, we will provide estimates for the velocity of the valence quarks just in the cases of charmonium, bottomonium and charm-bottom mesons, with $\ell=0$ and $\ell=1$. The results are displayed in \cref{tab:velocities}. These values are in line with the ones obtained by alternative potential models \cite{Domenech-Garret:2008itl} and are consistent with a nonrelativistic framework.

\begin{table}[h]
    \centering
    \begin{tabular}{|| c|c|c ||}
    \hline \hline
      Meson & $\ell$ & $\langle v^2 \rangle$ \\
      \hline \hline
       \multirow{2}{*}{Charmonium} & 0 & 0.26 \\
       \cline{2-3}
                                   & 1 & 0.32  \\
        \hline \hline
        \multirow{2}{*}{Charm-bottom mesons} & 0 & 0.10 \\
       \cline{2-3}
                                   & 1 & 0.13  \\
        \hline \hline
        \multirow{2}{*}{Bottomonium} & 0 & 0.05 \\
       \cline{2-3}
                                   & 1 & 0.06  \\
         \hline \hline
    \end{tabular}
    \caption{Squared velocity of valence quarks for charmonium, charm-bottom mesons and bottomonium.}
    \label{tab:velocities}
\end{table}

\section{Conclusions}\label{Section Discussion}

We evaluated the mass spectrum of charmonium, bottomonium and charm-bottom mesons by employing the CF model. As a first exploratory work in this framework, we restricted ourselves to the investigation of the spectrum of heavy mesons, since these are typically suited to be described within a nonrelativistic approach. We derived the Hamiltonian of the model by using the one-gluon exchange approximation. By adding to this Hamiltonian a confining Cornell potential we solved the Schrödinger equation for the ground state ($n=1$), restricting the analysis to states with angular momentum $\ell=0$ and $\ell=1$, where the nonrelativistic approximation is expected to be more reliable. We introduced corrections to this Hamiltonian, including spin-orbit terms, perturbatively. To fit the experimental/lattice data we explored the parameter space of the model, composed of the coupling, the gluon mass, the string tension and charm and bottom masses, by calculating the error between the results of the CF model and the experimental/lattice data. Whereas the latter three were treated as fixed quantities, we allowed the gauge coupling and the gluon mass to run with the energy scale, according to their ultraviolet behavior, computed in \cite{Pelaez:2014mxa}.

Our results display good agreement with experimental/lattice data, with the largest discrepancy being around 70 MeV. Additionally, the region of the parameter space that minimizes the error between the model and the data clearly favors the scenario with massive gluons, in line with other works, such as \cite{Parisi:1980jy,Mihara:2000wf,Field:2001iu,Fagundes:2011zx,Gutierrez-Guerrero:2019uwa,Paredes-Torres:2024mnz,Gutierrez-Guerrero:2024him}. However, as we plan to do in a future work, the investigation of other observables is needed in order to fully rule out the massless case. 

In general terms, our results show that the CF model is a very good tool to access the heavy meson spectra in the nonrelativistic limit. A more rigorous analysis of the accuracy of the results presented in this article requires a detailed study of the nonrelativistic limit of the BSE in the CF model, as we plan to do in a future work. 

\section{Acknowledgments} 
We would like to thank Nicolás Wschebor for useful discussions related to this work. We acknowledge the financial support from the Program for the Development of Basic Sciences (PEDECIBA), in particular, through its 'Despegue Científico' program, as well as from ANII and the Ministry of Education and Culture of Uruguay through the project FVF\_2023\_513. A. Alvez acknowledges the financial support from the Comisión Académica de Posgrado (CAP).

\appendix

\section{Fourier transform} \label{appendix_A_Fourier}

All the terms that appear in scattering amplitude \eqref{amplitud_charmbott} correspond to one of the following forms:
\begin{align}
    \textrm A: & \; \frac{1}{\mathbf q^2+m^2} & \textrm D: & 
    \; \frac{(\mathbf a
	\cdot \mathbf q )(\mathbf b\cdot \mathbf q)}{\mathbf q^2+m^2} \nonumber \\
    \textrm B: & \; \frac{\mathbf q}{\mathbf q^2+m^2} & \textrm 
    E: & \; \frac{(\mathbf a \cdot \mathbf q\,) (\mathbf b \cdot 
    \mathbf q)}{\mathbf q^2(\mathbf q^2+m^2)} 
    \nonumber \\
    \textrm C: & \; \frac{\mathbf q^2}{\mathbf q^2+m^2} & &
    \nonumber
\end{align}

For term (A) we calculate the Fourier transform by first integrating in the angular variables in spherical coordinates:

\begin{align}
	\int &\frac{d^3 q}{(2\pi)^3} \frac{e^{-i\mathbf q\cdot \mathbf
	r}}{\mathbf q^2+m^2} \nonumber \\
   & = \int\limits_0^\infty \frac{dq}{(2\pi)^3} 
	\frac{q^2}{q^2+m^2} \int d\Omega \sin\theta e^{-iqr\cos\theta}
	\nonumber \\ &= \frac{1}{ir(2\pi)^2}\int\limits_0^\infty dq \frac{q}
	{q^2+m^2}\left(e^{-iqr}-e^{+iqr}\right) \nonumber \\
	& = -\frac{1}{ir(2\pi)^2} \int\limits_{-\infty}^\infty dq \frac{q}
	{q^2+m^2}e^{iqr}
\end{align}
where  $\mathbf q$ is the module $|\mathbf q|=q$.
For the radial integral we use the residue theorem. We consider the upper half-plane which corresponds to the set of complex numbers with positive imaginary part where the only pole is $q = im$. Then, we have
\begin{align}
	\int \frac{d^3 q}{(2\pi)^3} \frac{e^{-i\mathbf q\cdot \mathbf
	r}}{\mathbf q^2+m^2} &= \frac{-1}{ir(2\pi)^2}\int\limits _{-\infty}
\nonumber	^\infty dq \frac{q}{q^2+m^2}e^{-iqr}
    \\ &= 4\pi^2 i \, \mathrm{Res}
	(q = im)
   \\ & = \frac{e^{-mr}}{4\pi r},
\end{align}
which corresponds to Yukawa potential. Note that terms (B) and (C) can be written as
derivatives of this potential with respect to $r$:

\begin{align}
	 \int\frac{d^3 q}{(2\pi)^3} \frac{\mathbf q \,e^{-i\mathbf
	 q\cdot \mathbf r}} {\mathbf q^2+m^2} &= i\nabla_r\int 
	 \frac{d^3 q}{(2\pi)^3} \frac{e^{-i\mathbf q\cdot 
    \mathbf r}} {\mathbf q^2+m^2}
    \nonumber \\ &= i\nabla_r \left(
    \frac{e^{-mr}}{4\pi r}\right)
    \nonumber \\ &=-i \left(m+ \frac{1}{r}\right) \frac{e^{-mr}}{4 \pi r^2} \mathbf{r},
\end{align}

\begin{align}
	 \int\frac{d^3 q}{(2\pi)^3} 
    \frac{\mathbf q^2\, e^{-i\mathbf{q} \cdot \mathbf{r}}} {q^2+m^2} &= -
	\nabla_r^2\int \frac{d^3 q}{(2\pi)^3} \frac{e^{-i\mathbf q\cdot 
    \mathbf r}} {\mathbf q^2+m^2} 
    \nonumber \\ &= - \nabla^2 \left(
    \frac{e^{-mr}}{4\pi r}\right)
    \nonumber \\ &= -\frac{m^2 e^{-mr}}{4 \pi r} + \delta^3(r).
\end{align}

For term D we have

\begin{align}
    \int &\frac{d^3 q}{(2\pi)^3}\frac{(\mathbf a \cdot \mathbf
    q)(\mathbf q \cdot \mathbf b)}
    {\mathbf q^2+m^2} e^{-i\mathbf q\cdot \mathbf r} 
    \\&= - (\mathbf a \cdot \nabla) (\mathbf b \cdot \nabla) \left( \frac{e^{-mr}}{4\pi r} \right) \nonumber \\
    &= \frac{(\mathbf a\cdot\mathbf b)}{3} \delta^3(r) \nonumber \\ 
    & \hspace{0.5cm}-\left\{  \frac{3(\mathbf a\cdot\mathbf r)(\mathbf b\cdot\mathbf r)}{r^5} \left( \frac{m^2r^2}{3} +mr+1 \right) \right. \nonumber \\
    & \hspace{0.5cm}- \left. \frac{\mathbf a\cdot \mathbf b}
    {r^3}(mr+1) \right \} \frac{e^{-mr}}{4\pi}.
\end{align}

Finally, we can rewrite the denominator of term (E) as

\begin{equation}
	\frac{1}{\mathbf q^2(\mathbf q^2 + m^2)} = \frac{1}{m^2 
	\mathbf q^2} - \frac{1}{m^2(\mathbf q^2 + m^2)},
\end{equation}
so we end up with a term identical to (D) and another similar one but with $m \rightarrow 0$. Taking this limit, we obtain

\begin{align}
 \int & \frac{d^3 q}{(2\pi)^3}\frac{(\mathbf a \cdot \mathbf q)(\mathbf q \cdot \mathbf b) } {\mathbf q^2(\mathbf q^2+m^2)}e^{-i\mathbf q\cdot \mathbf r}\\
   & = \frac{1}{4\pi} \left\{ \frac{\mathbf a\cdot \mathbf b}
    {r^3} - \frac{3(\mathbf a\cdot\mathbf r)(\mathbf b\cdot\mathbf r)
    }{r^5} \right\}  \nonumber \\ 
    & \hspace{0.5cm}+ \left\{ \frac{3(\mathbf a\cdot\mathbf r)(\mathbf b\cdot\mathbf r)}{r^5}\left(\frac{m^2r^2}{3}+mr+1\right) \right. \nonumber \\
    &\hspace{0.5cm} - \left. \frac{\mathbf a\cdot \mathbf b} {r^3}(mr+1) \right\} \frac{e^{-mr}}{4\pi}.
\end{align}

\section{Perturbative corrections} \label{app corrections}

In this section, we calculate the corrections, $T_i$, corresponding to the expectation values  of the terms $H_i$ that constitute the expression  \eqref{eq:pert_corr}.

Let us begin by calculating the correction to the kinetic term
\begin{equation}
	T_1 = \left\langle - \left( \frac{1}{M_1^3} + \frac{1}{M_2^3} \right)\frac{\mathbf p^4}{8} \right\rangle.
\end{equation}
which is diagonal in the basis ${| n, \ell, m, s_1, s_2, m_{1}, m_{2} \rangle }$ of energy eigenstates, orbital angular momentum, and spin operators of the two particles. We need to calculate the diagonal matrix elements in this basis.
From expresion \eqref{Ho}, we have

\begin{equation}
	\mathbf p^2 = 2\mu \left(H_0+\frac{g^2 e^{-mr}}{3\pi r}-b_0 r \right); \label{p2}
\end{equation}
hence,
\begin{align}
    T_1 =  -\frac{\mu^2}{2}& \left( \frac{1}{M_1^3} + \frac{1}{M_2^3} \right) \times \nonumber \\ 
    &\left\langle \biggl( H_0 + \frac{g^2 e^{-mr}}{3\pi r}-b_0 r \right) ^2 \biggr\rangle.
\end{align}

Since the terms we want to calculate depend only on $r$, and the spherical harmonics are normalized, we have that\\
\begin{align}
    T_1 = -\frac{4 \mu^2}{8} & \left( \frac{1}{M_1^3} + \frac{1}{M_2^3} \right)   \left \lbrace E_0^2 \int\limits_0 ^\infty |R_{n\ell} (r)|^2 r^2 dr\right . \nonumber \\
    & \left.+ E_{0}\frac{g^{2}}{3\pi}\int\limits_{0}^{\infty}e^{-mr}|R_{n\ell}(r)|^{2} r dr \right . \nonumber \\
    & -E_{0}b_{0}\int\limits_{0}^{\infty}|R_{n\ell}(r)|^{2}r^{3}dr
    \nonumber \\
    & +\frac{4g^{4}}{9\pi^{2}} \int\limits_{0}^{\infty}e^{-2mr} |R_{n\ell}(r)|^{2}dr
    \nonumber \\
    & -\frac{g^{2}b_{0}}{3\pi}\int\limits_{0}^{\infty}e^{-mr}|R_{n\ell}(r)|^{2}r^{2}dr
    \nonumber \\
    & \left.+b_{0}^{2}\int\limits_{0}^{\infty}|R_{n\ell}(r)|^{2}r^{4}dr \right \rbrace.
\end{align}

Next term, 
\begin{align}
T_{2}=\left\langle -\left(\frac{1}{M_{1}^{2}}+\frac{1}{M_{2}^{2}}\right)\frac{g^{2}m^{2}e^{-mr}}{24\pi r}\right\rangle ,
\end{align} 
is a function of $r$, so is diagonal in the same basis of states as the previous one. Its expectation value is 
\begin{align}
T_{2}=-\left(\frac{1}{M_{1}^{2}}+\frac{1}{M_{2}^{2}}\right)\frac{g^{2}m^{2}}{24\pi}\int\limits_{0}^{\infty}e^{-mr}|R_{n\ell}(r)|^{2}rdr.
\end{align}

By integrating the delta function against the modulus of the wave function, we obtain the third correction term, 
\begin{align}
T_{3}&=\left\langle \frac{g^{2}}{6}\left(\frac{1}{M_{1}^{2}}+\frac{1}{M_{2}^{2}}\right)\delta^{3}(r)\right\rangle \\
&=\frac{g^{2}}{6}\left(\frac{1}{M_{1}^{2}}+\frac{1}{M_{2}^{2}}\right)|\psi_{n\ell}(0)|^{2}.
\end{align}

Since the wave functions for states with $\ell \ne 0$ are zero at the origin, this correction only appears for states with $\ell=0$. In the ground state, we have
\begin{align}
T_{3}=\frac{g^{2}}{24\pi}\left(\frac{1}{M_{1}^{2}}+\frac{1}{M_{2}^{2}}\right)|R_{10}(0)|^{2}.
\end{align}

The spin-orbit term is
\begin{align}
T_{4}&=	g^{2}\left\langle (mr+1)\frac{e^{-mr}}{3\pi r^{3}}\left[\frac{1}{2M_{1}^{2}}+\frac{1}{M_{2}M_{1}}\right]\boldsymbol{s_{1}}\cdot\mathbf{L}\right. \nonumber \\
	& \left.+(mr+1)\frac{e^{-mr}}{3\pi r^{3}}\left[\frac{1}{2M_{2}^{2}}+\frac{1}{M_{2}M_{1}}\right]\boldsymbol{s_{2}}\cdot\mathbf{L}\right\rangle 
\end{align}

Let us define the spin-sum operator as $\mathbf{S}=\mathbf{s}_{1}+\mathbf{s}_{2}$. The operator $\mathbf{S}^{2}$ has eigenvalues $S(S+1)$, with $S={0,1}$. For states with $S=0$, the correction $T_{4}$ is zero. 

In the spin-sum basis, the operators $\boldsymbol{s_{1}}\cdot\mathbf{L}$ and $\boldsymbol{s_{2}}\cdot\mathbf{L}$ contribute equally. We will refer to either of the two operators as $\boldsymbol{s}\cdot\mathbf{L}$, such that $\boldsymbol{s}\cdot\mathbf{L}=\frac{\left(\mathbf{\mathbf{s}}_{1}+\mathbf{\mathbf{s}}_{2}\right)}{2}\cdot\mathbf{L}$.

We can also define the total angular momentum as $\mathbf{J}=\mathbf{S}+\mathbf{L}$, taking eigenvalues between $|\ell-S|$ y $\ell+S$. Note that for $S=1$ y $\ell=1$, $J=\{0,1,2\}$. We can write 
$$(\mathbf{s}_{1}+\mathbf{s}_{2})\cdot\mathbf{L}=\frac{1}{2}\left(\mathbf{J}^{2}-\mathbf{S}^{2}-\mathbf{L}^{2}\right).$$ This operator is diagonal in the basis of the sum of moments, and it has eigenvalues $J(J+1)-S(S+1)-\ell(\ell+1)$, that are zero if $\ell=0$. 

\begin{align}
T_{4}&=-\frac{g^{2}}{12\pi}\left(\left[\frac{2M_{1}+M_{2}}{M_{2}M_{1}^{2}}\right]+\left[\frac{2M_{2}+M_{1}}{M_{1}M_{2}^{2}}\right]\right) \times \nonumber \\
& \left(J(J+1)-4\right)\int\limits_{0}^{\infty}\frac{(mr+1)}{r}e^{-mr}|R_{21}(r)|^{2}dr 
\end{align}

Note that the previous terms, those which do not depend on spin, are also diagonal in the total angular momentum basis.

Now, let's consider 
\begin{align}
T_{5}=\left\langle \frac{g^{2}}{3\pi M_{1}M_{2}m^{2}}\left[\frac{\mathbf{p}^{2}}{r^{3}}-\frac{3(\mathbf{p}\cdot\mathbf{r})^{2}}{r^{5}}\right]\right\rangle. \label{T5}
\end{align}

In the first term of \eqref{T5} we can express $\mathbf{p}^{2}$ using $H_{0}$ as we did in \eqref{p2}. For the term in $(\mathbf{p}\cdot\mathbf{r})^{2}$ we can write the moment $\mathbf{p}$ as $\mathbf{p}=-i\nabla$ (considering that it should only act on the wave function). This leads to

\begin{align}
T_{5}=\frac{g^{2}}{3\pi M_{1}M_{2}m^{2}} & \left\{ 2\mu\left(E_{0}\int\limits_{0}^{\infty}\frac{|R_{n\ell}(r)|^{2}}{r}dr \right.\right. \nonumber \\
	& +\frac{g^{2}}{3\pi}\int\limits_{0}^{\infty}\frac{e^{-mr}}{r^{2}}|R_{n\ell}(r)|^{2}dr \nonumber \\
	& \left.-b_{0}\int\limits_{0}^{\infty}|R_{n\ell}(r)|^{2}dr\right)
	\nonumber \\ 
 & \left. +3\int\limits_{0}^{\infty}\frac{R_{n\ell}(r)}{r}R''_{n\ell}dr\right\} ,
\end{align}
where $R''_{n\ell}=\partial_{r}^{2}R_{n\ell}(r)$.
Analogously, the term 
\begin{align}
T_{6}=	&\left\langle -\frac{g^{2}e^{-mr}}{3\pi M_{1}M_{2}m^{2}}\left[\frac{\mathbf{p}^{2}}{r^{3}}(mr+1)\right.\right. \nonumber \\
	& \left.\left.-\frac{3(\mathbf{p}\cdot r)^{2}}{r^{5}}\left(\frac{m^{2}r^{2}}{3}+mr+1\right)\right]\right\rangle  
 \end{align}
 
takes the form 
 \begin{align}
 T_{6}&=	\frac{-g^{2}}{3\pi M_{1}M_{2}m^{2}} \times \nonumber \\ 
 &\left[ 2\mu\left(E_{0}\int\limits_{0}^{\infty}\frac{(mr+1)e^{-mr}}{r}|R_{n\ell}(r)|^{2}dr\right.\right. \nonumber \\
	&+\frac{g^{2}}{3\pi}\int\limits_{0}^{\infty}\frac{(mr+1)e^{-mr}}{r^{2}}|R_{n\ell}(r)|^{2}dr \nonumber \\
	& \left.-b_{0}\int\limits_{0}^{\infty}e^{-mr}(mr+1) |R_{n\ell}(r)|^{2}dr \right)  \\
	& \left.+\int\limits_{0}^{\infty}\frac{\left(m^{2}r^{2}+3mr+3\right)}{r}e^{-mr}R_{n\ell}R''_{n\ell}dr \right ]\nonumber
 \end{align}

The spin terms 
\begin{align}
T_{7}=\left\langle \frac{8}{9}\frac{g^{2}}{M_{1}M_{2}}(\boldsymbol{s}_{1}\cdot\boldsymbol{s}_{2})\delta^{3}(r)\right\rangle , 
\end{align}
and 
\begin{align}
T_{8}=\left\langle -\frac{g^{2}(\boldsymbol{s}_{1}\cdot\boldsymbol{s}_{2})e^{-mr}}{3\pi M_{1}M_{2}r^{3}}(m^{2}r^{2}+mr+1)\right\rangle ,
\end{align}
can be written in the spin sum basis considering $$\boldsymbol{s}_{1}\cdot\boldsymbol{s}_{2}=\frac{1}{2}\left[\mathbf{S}^{2}-\mathbf{s}_{1}^{2}-\mathbf{s}_{2}^{2}\right].$$ This operator is diagonal, with eigenvalues $\frac{1}{2}\left[S(S+1)-\frac{3}{2}\right]$, since the eigenvalues of both $\mathbf{s}_{1}^{2}$ and $\mathbf{s}_{2}^{2}$ are $\frac{3}{4}$. Then, in this basis 
\begin{align} 
T_{7}=\frac{g^{2}}{9\pi M_{1}M_{2}}\left[S(S+1)-\frac{3}{2}\right]|R_{n\ell}(0)|^{2},
\end{align}
and 
\begin{align}
T_{8}=& -\frac{g^{2}}{6\pi M_{1}M_{2}}\left[S(S+1)-\frac{3}{2}\right] \times \nonumber \\
&\int\limits_{0}^{\infty}dr\,\frac{e^{-\tilde{m}\rho}}{r}(m^{2}r^{2}+mr+1)|R_{n\ell}(r)|^{2}.
\end{align}
Finally, we have the term

\begin{align}
T_{9} = &  \left\langle  \frac{-g^{2} e^{-mr}}{3\pi M_{1}M_{2}} \frac{( \mathbf{s}_{1} \cdot \mathbf{r})(\mathbf{s}_{2}\cdot\mathbf{r})}{r^{5}} \times \right. \nonumber \\
&\hspace{1.7cm} \left. \left( \frac{m^{2}r^{2}}{3} + mr + 1 \right) \right\rangle
\end{align}

For two operators $\mathbf{a}$ y $\mathbf{b}$ that commute with each other and commute with the orbital angular momentum and with $\mathbf{r}$, it holds that 
\begin{align} 
\left\langle (\mathbf{a}\cdot\mathbf{b})r^{2}f(r)-3(\mathbf{a}\cdot\mathbf{r})\right.&\left.(\mathbf{b}\cdot\mathbf{r})f(r) \right \rangle = \nonumber \\
&- \frac{\left\langle r^{2}f(r) \right\rangle }{(2\ell+3)(2\ell-1)} \times \nonumber 
\\
&\left\langle 2\mathbf{L}^{2}(\mathbf{a}\cdot\mathbf{b})-3(\mathbf{a}\cdot\mathbf{L})(\mathbf{b}\cdot\mathbf{L}) \right. \nonumber
\\
&\left. -3(\mathbf{b}\cdot\mathbf{L})(\mathbf{a} \cdot\mathbf{L}) \right\rangle. \nonumber
\end{align}

We can apply this relation to the spin operators
 and express the products of the spins in terms of the total spin as $\mathbf{s}_{1}\cdot\mathbf{s}_{2}=\frac{1}{2}\left[\mathbf{S}^{2}-\mathbf{s}_{1}^{2}-\mathbf{s}_{2}^{2}\right]$ and the products $\mathbf{S}\cdot\mathbf{L}$ in terms of the sum of moments $\mathbf{J}=\mathbf{L}+\mathbf{S}$. This way, we obtain,
 \begin{align}
     \left\langle (\mathbf{s}_{1}\cdot\mathbf{s}_{2})r^{2}f(r)-3(\mathbf{s}_{1}\cdot\mathbf{r})\right.&\left.(\mathbf{s}_{2}\cdot\mathbf{r}) f(r)\right\rangle =\nonumber \\
     &\frac{-\left\langle r^{2}f(r)\right\rangle }{(2\ell+3)(2\ell-1)} \times \nonumber \\
     &\left\langle \mathbf{S}^{2} \mathbf{L}^{2}-\frac{3}{4}\left(\mathbf{J}^{2}-\mathbf{S}^{2}-\mathbf{L}^{2}\right) \right. \nonumber
     \\
    & \left.-\frac{3}{4}\left(\mathbf{J}^{2}-\mathbf{S}^{2}-\mathbf{L}^{2}\right)^{2}\right\rangle .\nonumber
\end{align}

The operator on the right-hand side of the last expression is diagonal in the total momentum basis. The mean value of the term on the left-hand side proportional to $(\boldsymbol{s}_{1}\cdot\mathbf{r})(\boldsymbol{s}_{2}\cdot\mathbf{r})$ takes the form
\begin{align}
 \left\langle 3(\mathbf{s}_{1}\cdot\mathbf{r})(\mathbf{s}_{2}\cdot\mathbf{r})f(r)\right\rangle& =	\frac{-\left\langle r^{2}f(r)\right\rangle }{(2\ell+3)(2\ell-1)} \left[ S(S+1)\ell(\ell+1) \right. \nonumber \\
& -\frac{3}{4}[J(J+1)-S(S+1)-\ell(\ell+1)]  \nonumber \\ 
& \left. -\frac{3}{4}[J(J+1)-S(S+1)-\ell(\ell+1)]^{2} \right] \nonumber \\
&-4\left[S(S+1)-\frac{3}{2}\right] \left \langle r^{2}f(r) \right \rangle . \nonumber
\end{align}
 
 We can finally write the correction $T_{9}$ as
 \begin{align}
 T_{9}&=\frac{g^{2}}{3\pi M_{1}M_{2}} \left \lbrace \frac{1}{(2\ell+3)(2\ell-1)} \biggl[ S(S+1)\ell(\ell+1) \right. \nonumber \\
&-\frac{3}{4}\left[J(J+1)-S(S+1)-\ell(\ell+1)\right] \nonumber \\
	&\left.- \frac{3}{4} \left[ J(J+1)- S(S+1)- \ell(\ell+1) \right]^{2} \right] \nonumber \\
 &\left. -4 \left[ S(S+1)-\frac{3}{2} \right] \right \rbrace \times \nonumber \\ &\int\limits_{0}^{\infty}dr\frac{e^{-mr}}{r} \left( \frac{m^{2}r^{2}}{3}+mr+1 \right) |R_{n\ell}(r)|^{2}.
\end{align}

\bibliography{main}

\begin{thebibliography}{106}%
\makeatletter
\providecommand \@ifxundefined [1]{%
 \@ifx{#1\undefined}
}%
\providecommand \@ifnum [1]{%
 \ifnum #1\expandafter \@firstoftwo
 \else \expandafter \@secondoftwo
 \fi
}%
\providecommand \@ifx [1]{%
 \ifx #1\expandafter \@firstoftwo
 \else \expandafter \@secondoftwo
 \fi
}%
\providecommand \natexlab [1]{#1}%
\providecommand \enquote  [1]{``#1''}%
\providecommand \bibnamefont  [1]{#1}%
\providecommand \bibfnamefont [1]{#1}%
\providecommand \citenamefont [1]{#1}%
\providecommand \href@noop [0]{\@secondoftwo}%
\providecommand \href [0]{\begingroup \@sanitize@url \@href}%
\providecommand \@href[1]{\@@startlink{#1}\@@href}%
\providecommand \@@href[1]{\endgroup#1\@@endlink}%
\providecommand \@sanitize@url [0]{\catcode `\\12\catcode `\$12\catcode
  `\&12\catcode `\#12\catcode `\^12\catcode `\_12\catcode `\%12\relax}%
\providecommand \@@startlink[1]{}%
\providecommand \@@endlink[0]{}%
\providecommand \url  [0]{\begingroup\@sanitize@url \@url }%
\providecommand \@url [1]{\endgroup\@href {#1}{\urlprefix }}%
\providecommand \urlprefix  [0]{URL }%
\providecommand \Eprint [0]{\href }%
\providecommand \doibase [0]{http://dx.doi.org/}%
\providecommand \selectlanguage [0]{\@gobble}%
\providecommand \bibinfo  [0]{\@secondoftwo}%
\providecommand \bibfield  [0]{\@secondoftwo}%
\providecommand \translation [1]{[#1]}%
\providecommand \BibitemOpen [0]{}%
\providecommand \bibitemStop [0]{}%
\providecommand \bibitemNoStop [0]{.\EOS\space}%
\providecommand \EOS [0]{\spacefactor3000\relax}%
\providecommand \BibitemShut  [1]{\csname bibitem#1\endcsname}%
\let\auto@bib@innerbib\@empty
\bibitem [{\citenamefont {Pel\'aez}\ \emph {et~al.}(2014)\citenamefont
  {Pel\'aez}, \citenamefont {Tissier},\ and\ \citenamefont
  {Wschebor}}]{Pelaez:2014mxa}%
  \BibitemOpen
  \bibfield  {author} {\bibinfo {author} {\bibfnamefont {M.}~\bibnamefont
  {Pel\'aez}}, \bibinfo {author} {\bibfnamefont {M.}~\bibnamefont {Tissier}}, \
  and\ \bibinfo {author} {\bibfnamefont {N.}~\bibnamefont {Wschebor}},\ }\href
  {\doibase 10.1103/PhysRevD.90.065031} {\bibfield  {journal} {\bibinfo
  {journal} {Phys. Rev. D}\ }\textbf {\bibinfo {volume} {90}},\ \bibinfo
  {pages} {065031} (\bibinfo {year} {2014})},\ \Eprint
  {http://arxiv.org/abs/1407.2005} {arXiv:1407.2005 [hep-th]} \BibitemShut
  {NoStop}%
\bibitem [{\citenamefont {Cucchieri}\ and\ \citenamefont
  {Mendes}(2008)}]{Cucchieri:2007rg}%
  \BibitemOpen
  \bibfield  {author} {\bibinfo {author} {\bibfnamefont {A.}~\bibnamefont
  {Cucchieri}}\ and\ \bibinfo {author} {\bibfnamefont {T.}~\bibnamefont
  {Mendes}},\ }\href {\doibase 10.1103/PhysRevLett.100.241601} {\bibfield
  {journal} {\bibinfo  {journal} {Phys. Rev. Lett.}\ }\textbf {\bibinfo
  {volume} {100}},\ \bibinfo {pages} {241601} (\bibinfo {year} {2008})},\
  \Eprint {http://arxiv.org/abs/0712.3517} {arXiv:0712.3517 [hep-lat]}
  \BibitemShut {NoStop}%
\bibitem [{\citenamefont {Bornyakov}\ \emph {et~al.}(2010)\citenamefont
  {Bornyakov}, \citenamefont {Mitrjushkin},\ and\ \citenamefont
  {Muller-Preussker}}]{Bornyakov:2009ug}%
  \BibitemOpen
  \bibfield  {author} {\bibinfo {author} {\bibfnamefont {V.~G.}\ \bibnamefont
  {Bornyakov}}, \bibinfo {author} {\bibfnamefont {V.~K.}\ \bibnamefont
  {Mitrjushkin}}, \ and\ \bibinfo {author} {\bibfnamefont {M.}~\bibnamefont
  {Muller-Preussker}},\ }\href {\doibase 10.1103/PhysRevD.81.054503} {\bibfield
   {journal} {\bibinfo  {journal} {Phys. Rev. D}\ }\textbf {\bibinfo {volume}
  {81}},\ \bibinfo {pages} {054503} (\bibinfo {year} {2010})},\ \Eprint
  {http://arxiv.org/abs/0912.4475} {arXiv:0912.4475 [hep-lat]} \BibitemShut
  {NoStop}%
\bibitem [{\citenamefont {Bogolubsky}\ \emph {et~al.}(2009)\citenamefont
  {Bogolubsky}, \citenamefont {Ilgenfritz}, \citenamefont {Muller-Preussker},\
  and\ \citenamefont {Sternbeck}}]{Bogolubsky:2009dc}%
  \BibitemOpen
  \bibfield  {author} {\bibinfo {author} {\bibfnamefont {I.~L.}\ \bibnamefont
  {Bogolubsky}}, \bibinfo {author} {\bibfnamefont {E.~M.}\ \bibnamefont
  {Ilgenfritz}}, \bibinfo {author} {\bibfnamefont {M.}~\bibnamefont
  {Muller-Preussker}}, \ and\ \bibinfo {author} {\bibfnamefont
  {A.}~\bibnamefont {Sternbeck}},\ }\href {\doibase
  10.1016/j.physletb.2009.04.076} {\bibfield  {journal} {\bibinfo  {journal}
  {Phys. Lett. B}\ }\textbf {\bibinfo {volume} {676}},\ \bibinfo {pages} {69}
  (\bibinfo {year} {2009})},\ \Eprint {http://arxiv.org/abs/0901.0736}
  {arXiv:0901.0736 [hep-lat]} \BibitemShut {NoStop}%
\bibitem [{\citenamefont {Maas}(2013)}]{Maas:2011se}%
  \BibitemOpen
  \bibfield  {author} {\bibinfo {author} {\bibfnamefont {A.}~\bibnamefont
  {Maas}},\ }\href {\doibase 10.1016/j.physrep.2012.11.002} {\bibfield
  {journal} {\bibinfo  {journal} {Phys. Rept.}\ }\textbf {\bibinfo {volume}
  {524}},\ \bibinfo {pages} {203} (\bibinfo {year} {2013})},\ \Eprint
  {http://arxiv.org/abs/1106.3942} {arXiv:1106.3942 [hep-ph]} \BibitemShut
  {NoStop}%
\bibitem [{\citenamefont {Oliveira}\ and\ \citenamefont
  {Silva}(2012)}]{Oliveira:2012eh}%
  \BibitemOpen
  \bibfield  {author} {\bibinfo {author} {\bibfnamefont {O.}~\bibnamefont
  {Oliveira}}\ and\ \bibinfo {author} {\bibfnamefont {P.~J.}\ \bibnamefont
  {Silva}},\ }\href {\doibase 10.1103/PhysRevD.86.114513} {\bibfield  {journal}
  {\bibinfo  {journal} {Phys. Rev. D}\ }\textbf {\bibinfo {volume} {86}},\
  \bibinfo {pages} {114513} (\bibinfo {year} {2012})},\ \Eprint
  {http://arxiv.org/abs/1207.3029} {arXiv:1207.3029 [hep-lat]} \BibitemShut
  {NoStop}%
\bibitem [{\citenamefont {Duarte}\ \emph {et~al.}(2016)\citenamefont {Duarte},
  \citenamefont {Oliveira},\ and\ \citenamefont {Silva}}]{Duarte:2016iko}%
  \BibitemOpen
  \bibfield  {author} {\bibinfo {author} {\bibfnamefont {A.~G.}\ \bibnamefont
  {Duarte}}, \bibinfo {author} {\bibfnamefont {O.}~\bibnamefont {Oliveira}}, \
  and\ \bibinfo {author} {\bibfnamefont {P.~J.}\ \bibnamefont {Silva}},\ }\href
  {\doibase 10.1103/PhysRevD.94.014502} {\bibfield  {journal} {\bibinfo
  {journal} {Phys. Rev. D}\ }\textbf {\bibinfo {volume} {94}},\ \bibinfo
  {pages} {014502} (\bibinfo {year} {2016})},\ \Eprint
  {http://arxiv.org/abs/1605.00594} {arXiv:1605.00594 [hep-lat]} \BibitemShut
  {NoStop}%
\bibitem [{\citenamefont {Aguilar}\ \emph {et~al.}(2008)\citenamefont
  {Aguilar}, \citenamefont {Binosi},\ and\ \citenamefont
  {Papavassiliou}}]{Aguilar:2008xm}%
  \BibitemOpen
  \bibfield  {author} {\bibinfo {author} {\bibfnamefont {A.~C.}\ \bibnamefont
  {Aguilar}}, \bibinfo {author} {\bibfnamefont {D.}~\bibnamefont {Binosi}}, \
  and\ \bibinfo {author} {\bibfnamefont {J.}~\bibnamefont {Papavassiliou}},\
  }\href {\doibase 10.1103/PhysRevD.78.025010} {\bibfield  {journal} {\bibinfo
  {journal} {Phys. Rev. D}\ }\textbf {\bibinfo {volume} {78}},\ \bibinfo
  {pages} {025010} (\bibinfo {year} {2008})},\ \Eprint
  {http://arxiv.org/abs/0802.1870} {arXiv:0802.1870 [hep-ph]} \BibitemShut
  {NoStop}%
\bibitem [{\citenamefont {Aguilar}\ \emph {et~al.}(2012)\citenamefont
  {Aguilar}, \citenamefont {Binosi},\ and\ \citenamefont
  {Papavassiliou}}]{Aguilar:2012rz}%
  \BibitemOpen
  \bibfield  {author} {\bibinfo {author} {\bibfnamefont {A.~C.}\ \bibnamefont
  {Aguilar}}, \bibinfo {author} {\bibfnamefont {D.}~\bibnamefont {Binosi}}, \
  and\ \bibinfo {author} {\bibfnamefont {J.}~\bibnamefont {Papavassiliou}},\
  }\href {\doibase 10.1103/PhysRevD.86.014032} {\bibfield  {journal} {\bibinfo
  {journal} {Phys. Rev. D}\ }\textbf {\bibinfo {volume} {86}},\ \bibinfo
  {pages} {014032} (\bibinfo {year} {2012})},\ \Eprint
  {http://arxiv.org/abs/1204.3868} {arXiv:1204.3868 [hep-ph]} \BibitemShut
  {NoStop}%
\bibitem [{\citenamefont {Huber}(2020{\natexlab{a}})}]{Huber:2018ned}%
  \BibitemOpen
  \bibfield  {author} {\bibinfo {author} {\bibfnamefont {M.~Q.}\ \bibnamefont
  {Huber}},\ }\href {\doibase 10.1016/j.physrep.2020.04.004} {\bibfield
  {journal} {\bibinfo  {journal} {Phys. Rept.}\ }\textbf {\bibinfo {volume}
  {879}},\ \bibinfo {pages} {1} (\bibinfo {year} {2020}{\natexlab{a}})},\
  \Eprint {http://arxiv.org/abs/1808.05227} {arXiv:1808.05227 [hep-ph]}
  \BibitemShut {NoStop}%
\bibitem [{\citenamefont {Huber}(2020{\natexlab{b}})}]{Huber:2020keu}%
  \BibitemOpen
  \bibfield  {author} {\bibinfo {author} {\bibfnamefont {M.~Q.}\ \bibnamefont
  {Huber}},\ }\href {\doibase 10.1103/PhysRevD.101.114009} {\bibfield
  {journal} {\bibinfo  {journal} {Phys. Rev. D}\ }\textbf {\bibinfo {volume}
  {101}},\ \bibinfo {pages} {114009} (\bibinfo {year} {2020}{\natexlab{b}})},\
  \Eprint {http://arxiv.org/abs/2003.13703} {arXiv:2003.13703 [hep-ph]}
  \BibitemShut {NoStop}%
\bibitem [{\citenamefont {Fischer}\ \emph {et~al.}(2009)\citenamefont
  {Fischer}, \citenamefont {Maas},\ and\ \citenamefont
  {Pawlowski}}]{Fischer:2008uz}%
  \BibitemOpen
  \bibfield  {author} {\bibinfo {author} {\bibfnamefont {C.~S.}\ \bibnamefont
  {Fischer}}, \bibinfo {author} {\bibfnamefont {A.}~\bibnamefont {Maas}}, \
  and\ \bibinfo {author} {\bibfnamefont {J.~M.}\ \bibnamefont {Pawlowski}},\
  }\href {\doibase 10.1016/j.aop.2009.07.009} {\bibfield  {journal} {\bibinfo
  {journal} {Annals Phys.}\ }\textbf {\bibinfo {volume} {324}},\ \bibinfo
  {pages} {2408} (\bibinfo {year} {2009})},\ \Eprint
  {http://arxiv.org/abs/0810.1987} {arXiv:0810.1987 [hep-ph]} \BibitemShut
  {NoStop}%
\bibitem [{\citenamefont {Cyrol}\ \emph {et~al.}(2016)\citenamefont {Cyrol},
  \citenamefont {Fister}, \citenamefont {Mitter}, \citenamefont {Pawlowski},\
  and\ \citenamefont {Strodthoff}}]{Cyrol:2016tym}%
  \BibitemOpen
  \bibfield  {author} {\bibinfo {author} {\bibfnamefont {A.~K.}\ \bibnamefont
  {Cyrol}}, \bibinfo {author} {\bibfnamefont {L.}~\bibnamefont {Fister}},
  \bibinfo {author} {\bibfnamefont {M.}~\bibnamefont {Mitter}}, \bibinfo
  {author} {\bibfnamefont {J.~M.}\ \bibnamefont {Pawlowski}}, \ and\ \bibinfo
  {author} {\bibfnamefont {N.}~\bibnamefont {Strodthoff}},\ }\href {\doibase
  10.1103/PhysRevD.94.054005} {\bibfield  {journal} {\bibinfo  {journal} {Phys.
  Rev. D}\ }\textbf {\bibinfo {volume} {94}},\ \bibinfo {pages} {054005}
  (\bibinfo {year} {2016})},\ \Eprint {http://arxiv.org/abs/1605.01856}
  {arXiv:1605.01856 [hep-ph]} \BibitemShut {NoStop}%
\bibitem [{\citenamefont {Cyrol}\ \emph {et~al.}(2018)\citenamefont {Cyrol},
  \citenamefont {Mitter}, \citenamefont {Pawlowski},\ and\ \citenamefont
  {Strodthoff}}]{Cyrol:2017ewj}%
  \BibitemOpen
  \bibfield  {author} {\bibinfo {author} {\bibfnamefont {A.~K.}\ \bibnamefont
  {Cyrol}}, \bibinfo {author} {\bibfnamefont {M.}~\bibnamefont {Mitter}},
  \bibinfo {author} {\bibfnamefont {J.~M.}\ \bibnamefont {Pawlowski}}, \ and\
  \bibinfo {author} {\bibfnamefont {N.}~\bibnamefont {Strodthoff}},\ }\href
  {\doibase 10.1103/PhysRevD.97.054006} {\bibfield  {journal} {\bibinfo
  {journal} {Phys. Rev. D}\ }\textbf {\bibinfo {volume} {97}},\ \bibinfo
  {pages} {054006} (\bibinfo {year} {2018})},\ \Eprint
  {http://arxiv.org/abs/1706.06326} {arXiv:1706.06326 [hep-ph]} \BibitemShut
  {NoStop}%
\bibitem [{\citenamefont {Dupuis}\ \emph {et~al.}(2021)\citenamefont {Dupuis},
  \citenamefont {Canet}, \citenamefont {Eichhorn}, \citenamefont {Metzner},
  \citenamefont {Pawlowski}, \citenamefont {Tissier},\ and\ \citenamefont
  {Wschebor}}]{Dupuis:2020fhh}%
  \BibitemOpen
  \bibfield  {author} {\bibinfo {author} {\bibfnamefont {N.}~\bibnamefont
  {Dupuis}}, \bibinfo {author} {\bibfnamefont {L.}~\bibnamefont {Canet}},
  \bibinfo {author} {\bibfnamefont {A.}~\bibnamefont {Eichhorn}}, \bibinfo
  {author} {\bibfnamefont {W.}~\bibnamefont {Metzner}}, \bibinfo {author}
  {\bibfnamefont {J.~M.}\ \bibnamefont {Pawlowski}}, \bibinfo {author}
  {\bibfnamefont {M.}~\bibnamefont {Tissier}}, \ and\ \bibinfo {author}
  {\bibfnamefont {N.}~\bibnamefont {Wschebor}},\ }\href {\doibase
  10.1016/j.physrep.2021.01.001} {\bibfield  {journal} {\bibinfo  {journal}
  {Phys. Rept.}\ }\textbf {\bibinfo {volume} {910}},\ \bibinfo {pages} {1}
  (\bibinfo {year} {2021})},\ \Eprint {http://arxiv.org/abs/2006.04853}
  {arXiv:2006.04853 [cond-mat.stat-mech]} \BibitemShut {NoStop}%
\bibitem [{\citenamefont {Aguilar}\ and\ \citenamefont
  {Natale}(2004)}]{Aguilar:2004sw}%
  \BibitemOpen
  \bibfield  {author} {\bibinfo {author} {\bibfnamefont {A.~C.}\ \bibnamefont
  {Aguilar}}\ and\ \bibinfo {author} {\bibfnamefont {A.~A.}\ \bibnamefont
  {Natale}},\ }\href {\doibase 10.1088/1126-6708/2004/08/057} {\bibfield
  {journal} {\bibinfo  {journal} {JHEP}\ }\textbf {\bibinfo {volume} {08}},\
  \bibinfo {pages} {057} (\bibinfo {year} {2004})},\ \Eprint
  {http://arxiv.org/abs/hep-ph/0408254} {arXiv:hep-ph/0408254} \BibitemShut
  {NoStop}%
\bibitem [{\citenamefont {Cucchieri}\ \emph {et~al.}(2008)\citenamefont
  {Cucchieri}, \citenamefont {Maas},\ and\ \citenamefont
  {Mendes}}]{Cucchieri:2008qm}%
  \BibitemOpen
  \bibfield  {author} {\bibinfo {author} {\bibfnamefont {A.}~\bibnamefont
  {Cucchieri}}, \bibinfo {author} {\bibfnamefont {A.}~\bibnamefont {Maas}}, \
  and\ \bibinfo {author} {\bibfnamefont {T.}~\bibnamefont {Mendes}},\ }\href
  {\doibase 10.1103/PhysRevD.77.094510} {\bibfield  {journal} {\bibinfo
  {journal} {Phys. Rev. D}\ }\textbf {\bibinfo {volume} {77}},\ \bibinfo
  {pages} {094510} (\bibinfo {year} {2008})},\ \Eprint
  {http://arxiv.org/abs/0803.1798} {arXiv:0803.1798 [hep-lat]} \BibitemShut
  {NoStop}%
\bibitem [{\citenamefont {Boucaud}\ \emph {et~al.}(2012)\citenamefont
  {Boucaud}, \citenamefont {Leroy}, \citenamefont {Yaouanc}, \citenamefont
  {Micheli}, \citenamefont {Pene},\ and\ \citenamefont
  {Rodriguez-Quintero}}]{Boucaud:2011ug}%
  \BibitemOpen
  \bibfield  {author} {\bibinfo {author} {\bibfnamefont {P.}~\bibnamefont
  {Boucaud}}, \bibinfo {author} {\bibfnamefont {J.~P.}\ \bibnamefont {Leroy}},
  \bibinfo {author} {\bibfnamefont {A.~L.}\ \bibnamefont {Yaouanc}}, \bibinfo
  {author} {\bibfnamefont {J.}~\bibnamefont {Micheli}}, \bibinfo {author}
  {\bibfnamefont {O.}~\bibnamefont {Pene}}, \ and\ \bibinfo {author}
  {\bibfnamefont {J.}~\bibnamefont {Rodriguez-Quintero}},\ }\href {\doibase
  10.1007/s00601-011-0301-2} {\bibfield  {journal} {\bibinfo  {journal} {Few
  Body Syst.}\ }\textbf {\bibinfo {volume} {53}},\ \bibinfo {pages} {387}
  (\bibinfo {year} {2012})},\ \Eprint {http://arxiv.org/abs/1109.1936}
  {arXiv:1109.1936 [hep-ph]} \BibitemShut {NoStop}%
\bibitem [{\citenamefont {Gribov}(1978)}]{Gribov:1977wm}%
  \BibitemOpen
  \bibfield  {author} {\bibinfo {author} {\bibfnamefont {V.~N.}\ \bibnamefont
  {Gribov}},\ }\href {\doibase 10.1016/0550-3213(78)90175-X} {\bibfield
  {journal} {\bibinfo  {journal} {Nucl. Phys. B}\ }\textbf {\bibinfo {volume}
  {139}},\ \bibinfo {pages} {1} (\bibinfo {year} {1978})}\BibitemShut {NoStop}%
\bibitem [{\citenamefont {Zwanziger}(1989)}]{Zwanziger:1989mf}%
  \BibitemOpen
  \bibfield  {author} {\bibinfo {author} {\bibfnamefont {D.}~\bibnamefont
  {Zwanziger}},\ }\href {\doibase 10.1016/0550-3213(89)90122-3} {\bibfield
  {journal} {\bibinfo  {journal} {Nucl. Phys. B}\ }\textbf {\bibinfo {volume}
  {323}},\ \bibinfo {pages} {513} (\bibinfo {year} {1989})}\BibitemShut
  {NoStop}%
\bibitem [{\citenamefont {Dudal}\ \emph {et~al.}(2008)\citenamefont {Dudal},
  \citenamefont {Gracey}, \citenamefont {Sorella}, \citenamefont
  {Vandersickel},\ and\ \citenamefont {Verschelde}}]{Dudal:2008sp}%
  \BibitemOpen
  \bibfield  {author} {\bibinfo {author} {\bibfnamefont {D.}~\bibnamefont
  {Dudal}}, \bibinfo {author} {\bibfnamefont {J.~A.}\ \bibnamefont {Gracey}},
  \bibinfo {author} {\bibfnamefont {S.~P.}\ \bibnamefont {Sorella}}, \bibinfo
  {author} {\bibfnamefont {N.}~\bibnamefont {Vandersickel}}, \ and\ \bibinfo
  {author} {\bibfnamefont {H.}~\bibnamefont {Verschelde}},\ }\href {\doibase
  10.1103/PhysRevD.78.065047} {\bibfield  {journal} {\bibinfo  {journal} {Phys.
  Rev. D}\ }\textbf {\bibinfo {volume} {78}},\ \bibinfo {pages} {065047}
  (\bibinfo {year} {2008})},\ \Eprint {http://arxiv.org/abs/0806.4348}
  {arXiv:0806.4348 [hep-th]} \BibitemShut {NoStop}%
\bibitem [{\citenamefont {Dudal}\ \emph {et~al.}(2010)\citenamefont {Dudal},
  \citenamefont {Oliveira},\ and\ \citenamefont {Vandersickel}}]{Dudal:2010tf}%
  \BibitemOpen
  \bibfield  {author} {\bibinfo {author} {\bibfnamefont {D.}~\bibnamefont
  {Dudal}}, \bibinfo {author} {\bibfnamefont {O.}~\bibnamefont {Oliveira}}, \
  and\ \bibinfo {author} {\bibfnamefont {N.}~\bibnamefont {Vandersickel}},\
  }\href {\doibase 10.1103/PhysRevD.81.074505} {\bibfield  {journal} {\bibinfo
  {journal} {Phys. Rev. D}\ }\textbf {\bibinfo {volume} {81}},\ \bibinfo
  {pages} {074505} (\bibinfo {year} {2010})},\ \Eprint
  {http://arxiv.org/abs/1002.2374} {arXiv:1002.2374 [hep-lat]} \BibitemShut
  {NoStop}%
\bibitem [{\citenamefont {Curci}\ and\ \citenamefont
  {Ferrari}(1976{\natexlab{a}})}]{Curci:1976bt}%
  \BibitemOpen
  \bibfield  {author} {\bibinfo {author} {\bibfnamefont {G.}~\bibnamefont
  {Curci}}\ and\ \bibinfo {author} {\bibfnamefont {R.}~\bibnamefont
  {Ferrari}},\ }\href {\doibase 10.1007/BF02729999} {\bibfield  {journal}
  {\bibinfo  {journal} {Nuovo Cim. A}\ }\textbf {\bibinfo {volume} {32}},\
  \bibinfo {pages} {151} (\bibinfo {year} {1976}{\natexlab{a}})}\BibitemShut
  {NoStop}%
\bibitem [{\citenamefont {Curci}\ and\ \citenamefont
  {Ferrari}(1976{\natexlab{b}})}]{Curci:1976kh}%
  \BibitemOpen
  \bibfield  {author} {\bibinfo {author} {\bibfnamefont {G.}~\bibnamefont
  {Curci}}\ and\ \bibinfo {author} {\bibfnamefont {R.}~\bibnamefont
  {Ferrari}},\ }\href {\doibase 10.1007/BF02730056} {\bibfield  {journal}
  {\bibinfo  {journal} {Nuovo Cim. A}\ }\textbf {\bibinfo {volume} {35}},\
  \bibinfo {pages} {1} (\bibinfo {year} {1976}{\natexlab{b}})},\ \bibinfo
  {note} {[Erratum: Nuovo Cim.A 47, 555 (1978)]}\BibitemShut {NoStop}%
\bibitem [{\citenamefont {Tissier}\ and\ \citenamefont
  {Wschebor}(2010)}]{Tissier:2010ts}%
  \BibitemOpen
  \bibfield  {author} {\bibinfo {author} {\bibfnamefont {M.}~\bibnamefont
  {Tissier}}\ and\ \bibinfo {author} {\bibfnamefont {N.}~\bibnamefont
  {Wschebor}},\ }\href {\doibase 10.1103/PhysRevD.82.101701} {\bibfield
  {journal} {\bibinfo  {journal} {Phys. Rev. D}\ }\textbf {\bibinfo {volume}
  {82}},\ \bibinfo {pages} {101701} (\bibinfo {year} {2010})},\ \Eprint
  {http://arxiv.org/abs/1004.1607} {arXiv:1004.1607 [hep-ph]} \BibitemShut
  {NoStop}%
\bibitem [{\citenamefont {Tissier}\ and\ \citenamefont
  {Wschebor}(2011)}]{Tissier:2011ey}%
  \BibitemOpen
  \bibfield  {author} {\bibinfo {author} {\bibfnamefont {M.}~\bibnamefont
  {Tissier}}\ and\ \bibinfo {author} {\bibfnamefont {N.}~\bibnamefont
  {Wschebor}},\ }\href {\doibase 10.1103/PhysRevD.84.045018} {\bibfield
  {journal} {\bibinfo  {journal} {Phys. Rev. D}\ }\textbf {\bibinfo {volume}
  {84}},\ \bibinfo {pages} {045018} (\bibinfo {year} {2011})},\ \Eprint
  {http://arxiv.org/abs/1105.2475} {arXiv:1105.2475 [hep-th]} \BibitemShut
  {NoStop}%
\bibitem [{\citenamefont {Reinosa}\ \emph {et~al.}(2017)\citenamefont
  {Reinosa}, \citenamefont {Serreau}, \citenamefont {Tissier},\ and\
  \citenamefont {Wschebor}}]{Reinosa:2017qtf}%
  \BibitemOpen
  \bibfield  {author} {\bibinfo {author} {\bibfnamefont {U.}~\bibnamefont
  {Reinosa}}, \bibinfo {author} {\bibfnamefont {J.}~\bibnamefont {Serreau}},
  \bibinfo {author} {\bibfnamefont {M.}~\bibnamefont {Tissier}}, \ and\
  \bibinfo {author} {\bibfnamefont {N.}~\bibnamefont {Wschebor}},\ }\href
  {\doibase 10.1103/PhysRevD.96.014005} {\bibfield  {journal} {\bibinfo
  {journal} {Phys. Rev. D}\ }\textbf {\bibinfo {volume} {96}},\ \bibinfo
  {pages} {014005} (\bibinfo {year} {2017})},\ \Eprint
  {http://arxiv.org/abs/1703.04041} {arXiv:1703.04041 [hep-th]} \BibitemShut
  {NoStop}%
\bibitem [{\citenamefont {Pelaez}\ \emph {et~al.}(2013)\citenamefont {Pelaez},
  \citenamefont {Tissier},\ and\ \citenamefont {Wschebor}}]{Pelaez:2013cpa}%
  \BibitemOpen
  \bibfield  {author} {\bibinfo {author} {\bibfnamefont {M.}~\bibnamefont
  {Pelaez}}, \bibinfo {author} {\bibfnamefont {M.}~\bibnamefont {Tissier}}, \
  and\ \bibinfo {author} {\bibfnamefont {N.}~\bibnamefont {Wschebor}},\ }\href
  {\doibase 10.1103/PhysRevD.88.125003} {\bibfield  {journal} {\bibinfo
  {journal} {Phys. Rev. D}\ }\textbf {\bibinfo {volume} {88}},\ \bibinfo
  {pages} {125003} (\bibinfo {year} {2013})},\ \Eprint
  {http://arxiv.org/abs/1310.2594} {arXiv:1310.2594 [hep-th]} \BibitemShut
  {NoStop}%
\bibitem [{\citenamefont {Pel\'aez}\ \emph {et~al.}(2015)\citenamefont
  {Pel\'aez}, \citenamefont {Tissier},\ and\ \citenamefont
  {Wschebor}}]{Pelaez:2015tba}%
  \BibitemOpen
  \bibfield  {author} {\bibinfo {author} {\bibfnamefont {M.}~\bibnamefont
  {Pel\'aez}}, \bibinfo {author} {\bibfnamefont {M.}~\bibnamefont {Tissier}}, \
  and\ \bibinfo {author} {\bibfnamefont {N.}~\bibnamefont {Wschebor}},\ }\href
  {\doibase 10.1103/PhysRevD.92.045012} {\bibfield  {journal} {\bibinfo
  {journal} {Phys. Rev. D}\ }\textbf {\bibinfo {volume} {92}},\ \bibinfo
  {pages} {045012} (\bibinfo {year} {2015})},\ \Eprint
  {http://arxiv.org/abs/1504.05157} {arXiv:1504.05157 [hep-th]} \BibitemShut
  {NoStop}%
\bibitem [{\citenamefont {Figueroa}\ and\ \citenamefont
  {Pel\'aez}(2022)}]{Figueroa:2021sjm}%
  \BibitemOpen
  \bibfield  {author} {\bibinfo {author} {\bibfnamefont {F.}~\bibnamefont
  {Figueroa}}\ and\ \bibinfo {author} {\bibfnamefont {M.}~\bibnamefont
  {Pel\'aez}},\ }\href {\doibase 10.1103/PhysRevD.105.094005} {\bibfield
  {journal} {\bibinfo  {journal} {Phys. Rev. D}\ }\textbf {\bibinfo {volume}
  {105}},\ \bibinfo {pages} {094005} (\bibinfo {year} {2022})},\ \Eprint
  {http://arxiv.org/abs/2110.09561} {arXiv:2110.09561 [hep-th]} \BibitemShut
  {NoStop}%
\bibitem [{\citenamefont {Barrios}\ \emph {et~al.}(2024)\citenamefont
  {Barrios}, \citenamefont {De~Fabritiis},\ and\ \citenamefont
  {Pel\'aez}}]{Barrios:2024ixj}%
  \BibitemOpen
  \bibfield  {author} {\bibinfo {author} {\bibfnamefont {N.}~\bibnamefont
  {Barrios}}, \bibinfo {author} {\bibfnamefont {P.}~\bibnamefont
  {De~Fabritiis}}, \ and\ \bibinfo {author} {\bibfnamefont {M.}~\bibnamefont
  {Pel\'aez}},\ }\href {\doibase 10.1103/PhysRevD.109.L091502} {\bibfield
  {journal} {\bibinfo  {journal} {Phys. Rev. D}\ }\textbf {\bibinfo {volume}
  {109}},\ \bibinfo {pages} {L091502} (\bibinfo {year} {2024})},\ \Eprint
  {http://arxiv.org/abs/2403.17056} {arXiv:2403.17056 [hep-th]} \BibitemShut
  {NoStop}%
\bibitem [{\citenamefont {Gracey}\ \emph {et~al.}(2019)\citenamefont {Gracey},
  \citenamefont {Pel\'aez}, \citenamefont {Reinosa},\ and\ \citenamefont
  {Tissier}}]{Gracey:2019xom}%
  \BibitemOpen
  \bibfield  {author} {\bibinfo {author} {\bibfnamefont {J.~A.}\ \bibnamefont
  {Gracey}}, \bibinfo {author} {\bibfnamefont {M.}~\bibnamefont {Pel\'aez}},
  \bibinfo {author} {\bibfnamefont {U.}~\bibnamefont {Reinosa}}, \ and\
  \bibinfo {author} {\bibfnamefont {M.}~\bibnamefont {Tissier}},\ }\href
  {\doibase 10.1103/PhysRevD.100.034023} {\bibfield  {journal} {\bibinfo
  {journal} {Phys. Rev. D}\ }\textbf {\bibinfo {volume} {100}},\ \bibinfo
  {pages} {034023} (\bibinfo {year} {2019})},\ \Eprint
  {http://arxiv.org/abs/1905.07262} {arXiv:1905.07262 [hep-th]} \BibitemShut
  {NoStop}%
\bibitem [{\citenamefont {Barrios}\ \emph {et~al.}(2020)\citenamefont
  {Barrios}, \citenamefont {Pel\'aez}, \citenamefont {Reinosa},\ and\
  \citenamefont {Wschebor}}]{Barrios:2020ubx}%
  \BibitemOpen
  \bibfield  {author} {\bibinfo {author} {\bibfnamefont {N.}~\bibnamefont
  {Barrios}}, \bibinfo {author} {\bibfnamefont {M.}~\bibnamefont {Pel\'aez}},
  \bibinfo {author} {\bibfnamefont {U.}~\bibnamefont {Reinosa}}, \ and\
  \bibinfo {author} {\bibfnamefont {N.}~\bibnamefont {Wschebor}},\ }\href
  {\doibase 10.1103/PhysRevD.102.114016} {\bibfield  {journal} {\bibinfo
  {journal} {Phys. Rev. D}\ }\textbf {\bibinfo {volume} {102}},\ \bibinfo
  {pages} {114016} (\bibinfo {year} {2020})},\ \Eprint
  {http://arxiv.org/abs/2009.00875} {arXiv:2009.00875 [hep-th]} \BibitemShut
  {NoStop}%
\bibitem [{\citenamefont {Barrios}\ \emph {et~al.}(2021)\citenamefont
  {Barrios}, \citenamefont {Gracey}, \citenamefont {Pel\'aez},\ and\
  \citenamefont {Reinosa}}]{Barrios:2021cks}%
  \BibitemOpen
  \bibfield  {author} {\bibinfo {author} {\bibfnamefont {N.}~\bibnamefont
  {Barrios}}, \bibinfo {author} {\bibfnamefont {J.~A.}\ \bibnamefont {Gracey}},
  \bibinfo {author} {\bibfnamefont {M.}~\bibnamefont {Pel\'aez}}, \ and\
  \bibinfo {author} {\bibfnamefont {U.}~\bibnamefont {Reinosa}},\ }\href
  {\doibase 10.1103/PhysRevD.104.094019} {\bibfield  {journal} {\bibinfo
  {journal} {Phys. Rev. D}\ }\textbf {\bibinfo {volume} {104}},\ \bibinfo
  {pages} {094019} (\bibinfo {year} {2021})},\ \Eprint
  {http://arxiv.org/abs/2103.16218} {arXiv:2103.16218 [hep-th]} \BibitemShut
  {NoStop}%
\bibitem [{\citenamefont {Barrios}\ \emph {et~al.}(2022)\citenamefont
  {Barrios}, \citenamefont {Pel\'aez},\ and\ \citenamefont
  {Reinosa}}]{Barrios:2022hzr}%
  \BibitemOpen
  \bibfield  {author} {\bibinfo {author} {\bibfnamefont {N.}~\bibnamefont
  {Barrios}}, \bibinfo {author} {\bibfnamefont {M.}~\bibnamefont {Pel\'aez}}, \
  and\ \bibinfo {author} {\bibfnamefont {U.}~\bibnamefont {Reinosa}},\ }\href
  {\doibase 10.1103/PhysRevD.106.114039} {\bibfield  {journal} {\bibinfo
  {journal} {Phys. Rev. D}\ }\textbf {\bibinfo {volume} {106}},\ \bibinfo
  {pages} {114039} (\bibinfo {year} {2022})},\ \Eprint
  {http://arxiv.org/abs/2207.10704} {arXiv:2207.10704 [hep-ph]} \BibitemShut
  {NoStop}%
\bibitem [{\citenamefont {Pel\'aez}\ \emph {et~al.}(2017)\citenamefont
  {Pel\'aez}, \citenamefont {Reinosa}, \citenamefont {Serreau}, \citenamefont
  {Tissier},\ and\ \citenamefont {Wschebor}}]{Pelaez:2017bhh}%
  \BibitemOpen
  \bibfield  {author} {\bibinfo {author} {\bibfnamefont {M.}~\bibnamefont
  {Pel\'aez}}, \bibinfo {author} {\bibfnamefont {U.}~\bibnamefont {Reinosa}},
  \bibinfo {author} {\bibfnamefont {J.}~\bibnamefont {Serreau}}, \bibinfo
  {author} {\bibfnamefont {M.}~\bibnamefont {Tissier}}, \ and\ \bibinfo
  {author} {\bibfnamefont {N.}~\bibnamefont {Wschebor}},\ }\href {\doibase
  10.1103/PhysRevD.96.114011} {\bibfield  {journal} {\bibinfo  {journal} {Phys.
  Rev. D}\ }\textbf {\bibinfo {volume} {96}},\ \bibinfo {pages} {114011}
  (\bibinfo {year} {2017})},\ \Eprint {http://arxiv.org/abs/1703.10288}
  {arXiv:1703.10288 [hep-th]} \BibitemShut {NoStop}%
\bibitem [{\citenamefont {Pel\'aez}\ \emph
  {et~al.}(2021{\natexlab{a}})\citenamefont {Pel\'aez}, \citenamefont
  {Reinosa}, \citenamefont {Serreau}, \citenamefont {Tissier},\ and\
  \citenamefont {Wschebor}}]{Pelaez:2020ups}%
  \BibitemOpen
  \bibfield  {author} {\bibinfo {author} {\bibfnamefont {M.}~\bibnamefont
  {Pel\'aez}}, \bibinfo {author} {\bibfnamefont {U.}~\bibnamefont {Reinosa}},
  \bibinfo {author} {\bibfnamefont {J.}~\bibnamefont {Serreau}}, \bibinfo
  {author} {\bibfnamefont {M.}~\bibnamefont {Tissier}}, \ and\ \bibinfo
  {author} {\bibfnamefont {N.}~\bibnamefont {Wschebor}},\ }\href {\doibase
  10.1103/PhysRevD.103.094035} {\bibfield  {journal} {\bibinfo  {journal}
  {Phys. Rev. D}\ }\textbf {\bibinfo {volume} {103}},\ \bibinfo {pages}
  {094035} (\bibinfo {year} {2021}{\natexlab{a}})},\ \Eprint
  {http://arxiv.org/abs/2010.13689} {arXiv:2010.13689 [hep-ph]} \BibitemShut
  {NoStop}%
\bibitem [{\citenamefont {Maelger}\ \emph
  {et~al.}(2018{\natexlab{a}})\citenamefont {Maelger}, \citenamefont
  {Reinosa},\ and\ \citenamefont {Serreau}}]{Maelger:2017amh}%
  \BibitemOpen
  \bibfield  {author} {\bibinfo {author} {\bibfnamefont {J.}~\bibnamefont
  {Maelger}}, \bibinfo {author} {\bibfnamefont {U.}~\bibnamefont {Reinosa}}, \
  and\ \bibinfo {author} {\bibfnamefont {J.}~\bibnamefont {Serreau}},\ }\href
  {\doibase 10.1103/PhysRevD.97.074027} {\bibfield  {journal} {\bibinfo
  {journal} {Phys. Rev. D}\ }\textbf {\bibinfo {volume} {97}},\ \bibinfo
  {pages} {074027} (\bibinfo {year} {2018}{\natexlab{a}})},\ \Eprint
  {http://arxiv.org/abs/1710.01930} {arXiv:1710.01930 [hep-ph]} \BibitemShut
  {NoStop}%
\bibitem [{\citenamefont {Maelger}\ \emph
  {et~al.}(2018{\natexlab{b}})\citenamefont {Maelger}, \citenamefont
  {Reinosa},\ and\ \citenamefont {Serreau}}]{Maelger:2018vow}%
  \BibitemOpen
  \bibfield  {author} {\bibinfo {author} {\bibfnamefont {J.}~\bibnamefont
  {Maelger}}, \bibinfo {author} {\bibfnamefont {U.}~\bibnamefont {Reinosa}}, \
  and\ \bibinfo {author} {\bibfnamefont {J.}~\bibnamefont {Serreau}},\ }\href
  {\doibase 10.1103/PhysRevD.98.094020} {\bibfield  {journal} {\bibinfo
  {journal} {Phys. Rev. D}\ }\textbf {\bibinfo {volume} {98}},\ \bibinfo
  {pages} {094020} (\bibinfo {year} {2018}{\natexlab{b}})},\ \Eprint
  {http://arxiv.org/abs/1805.10015} {arXiv:1805.10015 [hep-th]} \BibitemShut
  {NoStop}%
\bibitem [{\citenamefont {Maelger}\ \emph {et~al.}(2020)\citenamefont
  {Maelger}, \citenamefont {Reinosa},\ and\ \citenamefont
  {Serreau}}]{Maelger:2019cbk}%
  \BibitemOpen
  \bibfield  {author} {\bibinfo {author} {\bibfnamefont {J.}~\bibnamefont
  {Maelger}}, \bibinfo {author} {\bibfnamefont {U.}~\bibnamefont {Reinosa}}, \
  and\ \bibinfo {author} {\bibfnamefont {J.}~\bibnamefont {Serreau}},\ }\href
  {\doibase 10.1103/PhysRevD.101.014028} {\bibfield  {journal} {\bibinfo
  {journal} {Phys. Rev. D}\ }\textbf {\bibinfo {volume} {101}},\ \bibinfo
  {pages} {014028} (\bibinfo {year} {2020})},\ \Eprint
  {http://arxiv.org/abs/1903.04184} {arXiv:1903.04184 [hep-th]} \BibitemShut
  {NoStop}%
\bibitem [{\citenamefont {Suenaga}\ and\ \citenamefont
  {Kojo}(2019)}]{Suenaga:2019jjv}%
  \BibitemOpen
  \bibfield  {author} {\bibinfo {author} {\bibfnamefont {D.}~\bibnamefont
  {Suenaga}}\ and\ \bibinfo {author} {\bibfnamefont {T.}~\bibnamefont {Kojo}},\
  }\href {\doibase 10.1103/PhysRevD.100.076017} {\bibfield  {journal} {\bibinfo
   {journal} {Phys. Rev. D}\ }\textbf {\bibinfo {volume} {100}},\ \bibinfo
  {pages} {076017} (\bibinfo {year} {2019})},\ \Eprint
  {http://arxiv.org/abs/1905.08751} {arXiv:1905.08751 [hep-ph]} \BibitemShut
  {NoStop}%
\bibitem [{\citenamefont {Reinosa}(2019)}]{Reinosa:2019xqq}%
  \BibitemOpen
  \bibfield  {author} {\bibinfo {author} {\bibfnamefont {U.}~\bibnamefont
  {Reinosa}},\ }\href {\doibase 10.1007/978-3-031-11375-8} {\  (\bibinfo {year}
  {2019}),\ 10.1007/978-3-031-11375-8},\ \Eprint
  {http://arxiv.org/abs/2009.04933} {arXiv:2009.04933 [hep-th]} \BibitemShut
  {NoStop}%
\bibitem [{\citenamefont {Song}\ \emph {et~al.}(2019)\citenamefont {Song},
  \citenamefont {Baym}, \citenamefont {Hatsuda},\ and\ \citenamefont
  {Kojo}}]{Song:2019qoh}%
  \BibitemOpen
  \bibfield  {author} {\bibinfo {author} {\bibfnamefont {Y.}~\bibnamefont
  {Song}}, \bibinfo {author} {\bibfnamefont {G.}~\bibnamefont {Baym}}, \bibinfo
  {author} {\bibfnamefont {T.}~\bibnamefont {Hatsuda}}, \ and\ \bibinfo
  {author} {\bibfnamefont {T.}~\bibnamefont {Kojo}},\ }\href {\doibase
  10.1103/PhysRevD.100.034018} {\bibfield  {journal} {\bibinfo  {journal}
  {Phys. Rev. D}\ }\textbf {\bibinfo {volume} {100}},\ \bibinfo {pages}
  {034018} (\bibinfo {year} {2019})},\ \Eprint
  {http://arxiv.org/abs/1905.01005} {arXiv:1905.01005 [astro-ph.HE]}
  \BibitemShut {NoStop}%
\bibitem [{\citenamefont {van Egmond}\ \emph {et~al.}(2022)\citenamefont {van
  Egmond}, \citenamefont {Reinosa}, \citenamefont {Serreau},\ and\
  \citenamefont {Tissier}}]{vanEgmond:2021jyx}%
  \BibitemOpen
  \bibfield  {author} {\bibinfo {author} {\bibfnamefont {D.~M.}\ \bibnamefont
  {van Egmond}}, \bibinfo {author} {\bibfnamefont {U.}~\bibnamefont {Reinosa}},
  \bibinfo {author} {\bibfnamefont {J.}~\bibnamefont {Serreau}}, \ and\
  \bibinfo {author} {\bibfnamefont {M.}~\bibnamefont {Tissier}},\ }\href
  {\doibase 10.21468/SciPostPhys.12.3.087} {\bibfield  {journal} {\bibinfo
  {journal} {SciPost Phys.}\ }\textbf {\bibinfo {volume} {12}},\ \bibinfo
  {pages} {087} (\bibinfo {year} {2022})},\ \Eprint
  {http://arxiv.org/abs/2104.08974} {arXiv:2104.08974 [hep-ph]} \BibitemShut
  {NoStop}%
\bibitem [{\citenamefont {Surkau}\ and\ \citenamefont
  {Reinosa}(2024)}]{Surkau:2024zfb}%
  \BibitemOpen
  \bibfield  {author} {\bibinfo {author} {\bibfnamefont {V.~T.~M.}\
  \bibnamefont {Surkau}}\ and\ \bibinfo {author} {\bibfnamefont
  {U.}~\bibnamefont {Reinosa}},\ }\href@noop {} {\  (\bibinfo {year} {2024})},\
  \Eprint {http://arxiv.org/abs/2401.17869} {arXiv:2401.17869 [hep-ph]}
  \BibitemShut {NoStop}%
\bibitem [{\citenamefont {Oribe}\ \emph {et~al.}(2025)\citenamefont {Oribe},
  \citenamefont {Pel\'aez},\ and\ \citenamefont {Reinosa}}]{Oribe:2025ezp}%
  \BibitemOpen
  \bibfield  {author} {\bibinfo {author} {\bibfnamefont {S.}~\bibnamefont
  {Oribe}}, \bibinfo {author} {\bibfnamefont {M.}~\bibnamefont {Pel\'aez}}, \
  and\ \bibinfo {author} {\bibfnamefont {U.}~\bibnamefont {Reinosa}},\
  }\href@noop {} {\  (\bibinfo {year} {2025})},\ \Eprint
  {http://arxiv.org/abs/2503.14291} {arXiv:2503.14291 [hep-ph]} \BibitemShut
  {NoStop}%
\bibitem [{\citenamefont {Pel\'aez}\ \emph {et~al.}(2024)\citenamefont
  {Pel\'aez}, \citenamefont {Reinosa}, \citenamefont {Serreau}, \citenamefont
  {Tissier},\ and\ \citenamefont {Wschebor}}]{Pelaez:2024mtq}%
  \BibitemOpen
  \bibfield  {author} {\bibinfo {author} {\bibfnamefont {M.}~\bibnamefont
  {Pel\'aez}}, \bibinfo {author} {\bibfnamefont {U.}~\bibnamefont {Reinosa}},
  \bibinfo {author} {\bibfnamefont {J.}~\bibnamefont {Serreau}}, \bibinfo
  {author} {\bibfnamefont {M.}~\bibnamefont {Tissier}}, \ and\ \bibinfo
  {author} {\bibfnamefont {N.}~\bibnamefont {Wschebor}},\ }\href@noop {} {\
  (\bibinfo {year} {2024})},\ \Eprint {http://arxiv.org/abs/2405.20532}
  {arXiv:2405.20532 [hep-th]} \BibitemShut {NoStop}%
\bibitem [{\citenamefont {Bopsin}\ \emph {et~al.}(2025)\citenamefont {Bopsin},
  \citenamefont {El-Bennich}, \citenamefont {Krein}, \citenamefont {Serna},\
  and\ \citenamefont {da~Silveira}}]{Bopsin:2025vhz}%
  \BibitemOpen
  \bibfield  {author} {\bibinfo {author} {\bibfnamefont {G.~B.}\ \bibnamefont
  {Bopsin}}, \bibinfo {author} {\bibfnamefont {B.}~\bibnamefont {El-Bennich}},
  \bibinfo {author} {\bibfnamefont {G.}~\bibnamefont {Krein}}, \bibinfo
  {author} {\bibfnamefont {F.~E.}\ \bibnamefont {Serna}}, \ and\ \bibinfo
  {author} {\bibfnamefont {R.~C.}\ \bibnamefont {da~Silveira}},\ }\href@noop {}
  {\  (\bibinfo {year} {2025})},\ \Eprint {http://arxiv.org/abs/2507.12544}
  {arXiv:2507.12544 [hep-ph]} \BibitemShut {NoStop}%
\bibitem [{\citenamefont {Pel\'aez}\ \emph
  {et~al.}(2021{\natexlab{b}})\citenamefont {Pel\'aez}, \citenamefont
  {Reinosa}, \citenamefont {Serreau}, \citenamefont {Tissier},\ and\
  \citenamefont {Wschebor}}]{Pelaez:2021tpq}%
  \BibitemOpen
  \bibfield  {author} {\bibinfo {author} {\bibfnamefont {M.}~\bibnamefont
  {Pel\'aez}}, \bibinfo {author} {\bibfnamefont {U.}~\bibnamefont {Reinosa}},
  \bibinfo {author} {\bibfnamefont {J.}~\bibnamefont {Serreau}}, \bibinfo
  {author} {\bibfnamefont {M.}~\bibnamefont {Tissier}}, \ and\ \bibinfo
  {author} {\bibfnamefont {N.}~\bibnamefont {Wschebor}},\ }\href {\doibase
  10.1088/1361-6633/ac36b8} {\bibfield  {journal} {\bibinfo  {journal} {Rept.
  Prog. Phys.}\ }\textbf {\bibinfo {volume} {84}},\ \bibinfo {pages} {124202}
  (\bibinfo {year} {2021}{\natexlab{b}})},\ \Eprint
  {http://arxiv.org/abs/2106.04526} {arXiv:2106.04526 [hep-th]} \BibitemShut
  {NoStop}%
\bibitem [{\citenamefont {Brambilla}\ \emph {et~al.}(2014)\citenamefont
  {Brambilla} \emph {et~al.}}]{Brambilla:2014jmp}%
  \BibitemOpen
  \bibfield  {author} {\bibinfo {author} {\bibfnamefont {N.}~\bibnamefont
  {Brambilla}} \emph {et~al.},\ }\href {\doibase
  10.1140/epjc/s10052-014-2981-5} {\bibfield  {journal} {\bibinfo  {journal}
  {Eur. Phys. J. C}\ }\textbf {\bibinfo {volume} {74}},\ \bibinfo {pages}
  {2981} (\bibinfo {year} {2014})},\ \Eprint {http://arxiv.org/abs/1404.3723}
  {arXiv:1404.3723 [hep-ph]} \BibitemShut {NoStop}%
\bibitem [{\citenamefont {Caswell}\ and\ \citenamefont
  {Lepage}(1986)}]{Caswell:1985ui}%
  \BibitemOpen
  \bibfield  {author} {\bibinfo {author} {\bibfnamefont {W.~E.}\ \bibnamefont
  {Caswell}}\ and\ \bibinfo {author} {\bibfnamefont {G.~P.}\ \bibnamefont
  {Lepage}},\ }\href {\doibase 10.1016/0370-2693(86)91297-9} {\bibfield
  {journal} {\bibinfo  {journal} {Phys. Lett. B}\ }\textbf {\bibinfo {volume}
  {167}},\ \bibinfo {pages} {437} (\bibinfo {year} {1986})}\BibitemShut
  {NoStop}%
\bibitem [{\citenamefont {Bodwin}\ \emph {et~al.}(1995)\citenamefont {Bodwin},
  \citenamefont {Braaten},\ and\ \citenamefont {Lepage}}]{Bodwin:1994jh}%
  \BibitemOpen
  \bibfield  {author} {\bibinfo {author} {\bibfnamefont {G.~T.}\ \bibnamefont
  {Bodwin}}, \bibinfo {author} {\bibfnamefont {E.}~\bibnamefont {Braaten}}, \
  and\ \bibinfo {author} {\bibfnamefont {G.~P.}\ \bibnamefont {Lepage}},\
  }\href {\doibase 10.1103/PhysRevD.51.1125} {\bibfield  {journal} {\bibinfo
  {journal} {Phys. Rev. D}\ }\textbf {\bibinfo {volume} {51}},\ \bibinfo
  {pages} {1125} (\bibinfo {year} {1995})},\ \bibinfo {note} {[Erratum:
  Phys.Rev.D 55, 5853 (1997)]},\ \Eprint {http://arxiv.org/abs/hep-ph/9407339}
  {arXiv:hep-ph/9407339} \BibitemShut {NoStop}%
\bibitem [{\citenamefont {Pineda}\ and\ \citenamefont
  {Soto}(1998)}]{Pineda:1997bj}%
  \BibitemOpen
  \bibfield  {author} {\bibinfo {author} {\bibfnamefont {A.}~\bibnamefont
  {Pineda}}\ and\ \bibinfo {author} {\bibfnamefont {J.}~\bibnamefont {Soto}},\
  }\href {\doibase 10.1016/S0920-5632(97)01102-X} {\bibfield  {journal}
  {\bibinfo  {journal} {Nucl. Phys. B Proc. Suppl.}\ }\textbf {\bibinfo
  {volume} {64}},\ \bibinfo {pages} {428} (\bibinfo {year} {1998})},\ \Eprint
  {http://arxiv.org/abs/hep-ph/9707481} {arXiv:hep-ph/9707481} \BibitemShut
  {NoStop}%
\bibitem [{\citenamefont {Brambilla}\ \emph {et~al.}(2000)\citenamefont
  {Brambilla}, \citenamefont {Pineda}, \citenamefont {Soto},\ and\
  \citenamefont {Vairo}}]{Brambilla:1999xf}%
  \BibitemOpen
  \bibfield  {author} {\bibinfo {author} {\bibfnamefont {N.}~\bibnamefont
  {Brambilla}}, \bibinfo {author} {\bibfnamefont {A.}~\bibnamefont {Pineda}},
  \bibinfo {author} {\bibfnamefont {J.}~\bibnamefont {Soto}}, \ and\ \bibinfo
  {author} {\bibfnamefont {A.}~\bibnamefont {Vairo}},\ }\href {\doibase
  10.1016/S0550-3213(99)00693-8} {\bibfield  {journal} {\bibinfo  {journal}
  {Nucl. Phys. B}\ }\textbf {\bibinfo {volume} {566}},\ \bibinfo {pages} {275}
  (\bibinfo {year} {2000})},\ \Eprint {http://arxiv.org/abs/hep-ph/9907240}
  {arXiv:hep-ph/9907240} \BibitemShut {NoStop}%
\bibitem [{\citenamefont {Brambilla}\ \emph {et~al.}(2005)\citenamefont
  {Brambilla}, \citenamefont {Pineda}, \citenamefont {Soto},\ and\
  \citenamefont {Vairo}}]{Brambilla:2004jw}%
  \BibitemOpen
  \bibfield  {author} {\bibinfo {author} {\bibfnamefont {N.}~\bibnamefont
  {Brambilla}}, \bibinfo {author} {\bibfnamefont {A.}~\bibnamefont {Pineda}},
  \bibinfo {author} {\bibfnamefont {J.}~\bibnamefont {Soto}}, \ and\ \bibinfo
  {author} {\bibfnamefont {A.}~\bibnamefont {Vairo}},\ }\href {\doibase
  10.1103/RevModPhys.77.1423} {\bibfield  {journal} {\bibinfo  {journal} {Rev.
  Mod. Phys.}\ }\textbf {\bibinfo {volume} {77}},\ \bibinfo {pages} {1423}
  (\bibinfo {year} {2005})},\ \Eprint {http://arxiv.org/abs/hep-ph/0410047}
  {arXiv:hep-ph/0410047} \BibitemShut {NoStop}%
\bibitem [{\citenamefont {Dudek}\ \emph {et~al.}(2008)\citenamefont {Dudek},
  \citenamefont {Edwards}, \citenamefont {Mathur},\ and\ \citenamefont
  {Richards}}]{Dudek:2007wv}%
  \BibitemOpen
  \bibfield  {author} {\bibinfo {author} {\bibfnamefont {J.~J.}\ \bibnamefont
  {Dudek}}, \bibinfo {author} {\bibfnamefont {R.~G.}\ \bibnamefont {Edwards}},
  \bibinfo {author} {\bibfnamefont {N.}~\bibnamefont {Mathur}}, \ and\ \bibinfo
  {author} {\bibfnamefont {D.~G.}\ \bibnamefont {Richards}},\ }\href {\doibase
  10.1103/PhysRevD.77.034501} {\bibfield  {journal} {\bibinfo  {journal} {Phys.
  Rev. D}\ }\textbf {\bibinfo {volume} {77}},\ \bibinfo {pages} {034501}
  (\bibinfo {year} {2008})},\ \Eprint {http://arxiv.org/abs/0707.4162}
  {arXiv:0707.4162 [hep-lat]} \BibitemShut {NoStop}%
\bibitem [{\citenamefont {Thacker}\ and\ \citenamefont
  {Lepage}(1991)}]{Thacker:1990bm}%
  \BibitemOpen
  \bibfield  {author} {\bibinfo {author} {\bibfnamefont {B.~A.}\ \bibnamefont
  {Thacker}}\ and\ \bibinfo {author} {\bibfnamefont {G.~P.}\ \bibnamefont
  {Lepage}},\ }\href {\doibase 10.1103/PhysRevD.43.196} {\bibfield  {journal}
  {\bibinfo  {journal} {Phys. Rev. D}\ }\textbf {\bibinfo {volume} {43}},\
  \bibinfo {pages} {196} (\bibinfo {year} {1991})}\BibitemShut {NoStop}%
\bibitem [{\citenamefont {Lepage}\ \emph {et~al.}(1992)\citenamefont {Lepage},
  \citenamefont {Magnea}, \citenamefont {Nakhleh}, \citenamefont {Magnea},\
  and\ \citenamefont {Hornbostel}}]{Lepage:1992tx}%
  \BibitemOpen
  \bibfield  {author} {\bibinfo {author} {\bibfnamefont {G.~P.}\ \bibnamefont
  {Lepage}}, \bibinfo {author} {\bibfnamefont {L.}~\bibnamefont {Magnea}},
  \bibinfo {author} {\bibfnamefont {C.}~\bibnamefont {Nakhleh}}, \bibinfo
  {author} {\bibfnamefont {U.}~\bibnamefont {Magnea}}, \ and\ \bibinfo {author}
  {\bibfnamefont {K.}~\bibnamefont {Hornbostel}},\ }\href {\doibase
  10.1103/PhysRevD.46.4052} {\bibfield  {journal} {\bibinfo  {journal} {Phys.
  Rev. D}\ }\textbf {\bibinfo {volume} {46}},\ \bibinfo {pages} {4052}
  (\bibinfo {year} {1992})},\ \Eprint {http://arxiv.org/abs/hep-lat/9205007}
  {arXiv:hep-lat/9205007} \BibitemShut {NoStop}%
\bibitem [{\citenamefont {Blank}\ and\ \citenamefont
  {Krassnigg}(2011)}]{Blank:2011ha}%
  \BibitemOpen
  \bibfield  {author} {\bibinfo {author} {\bibfnamefont {M.}~\bibnamefont
  {Blank}}\ and\ \bibinfo {author} {\bibfnamefont {A.}~\bibnamefont
  {Krassnigg}},\ }\href {\doibase 10.1103/PhysRevD.84.096014} {\bibfield
  {journal} {\bibinfo  {journal} {Phys. Rev. D}\ }\textbf {\bibinfo {volume}
  {84}},\ \bibinfo {pages} {096014} (\bibinfo {year} {2011})},\ \Eprint
  {http://arxiv.org/abs/1109.6509} {arXiv:1109.6509 [hep-ph]} \BibitemShut
  {NoStop}%
\bibitem [{\citenamefont {Fischer}\ \emph {et~al.}(2015)\citenamefont
  {Fischer}, \citenamefont {Kubrak},\ and\ \citenamefont
  {Williams}}]{Fischer:2014cfa}%
  \BibitemOpen
  \bibfield  {author} {\bibinfo {author} {\bibfnamefont {C.~S.}\ \bibnamefont
  {Fischer}}, \bibinfo {author} {\bibfnamefont {S.}~\bibnamefont {Kubrak}}, \
  and\ \bibinfo {author} {\bibfnamefont {R.}~\bibnamefont {Williams}},\ }\href
  {\doibase 10.1140/epja/i2015-15010-7} {\bibfield  {journal} {\bibinfo
  {journal} {Eur. Phys. J. A}\ }\textbf {\bibinfo {volume} {51}},\ \bibinfo
  {pages} {10} (\bibinfo {year} {2015})},\ \Eprint
  {http://arxiv.org/abs/1409.5076} {arXiv:1409.5076 [hep-ph]} \BibitemShut
  {NoStop}%
\bibitem [{\citenamefont {Hilger}\ \emph {et~al.}(2015)\citenamefont {Hilger},
  \citenamefont {Popovici}, \citenamefont {Gomez-Rocha},\ and\ \citenamefont
  {Krassnigg}}]{Hilger:2014nma}%
  \BibitemOpen
  \bibfield  {author} {\bibinfo {author} {\bibfnamefont {T.}~\bibnamefont
  {Hilger}}, \bibinfo {author} {\bibfnamefont {C.}~\bibnamefont {Popovici}},
  \bibinfo {author} {\bibfnamefont {M.}~\bibnamefont {Gomez-Rocha}}, \ and\
  \bibinfo {author} {\bibfnamefont {A.}~\bibnamefont {Krassnigg}},\ }\href
  {\doibase 10.1103/PhysRevD.91.034013} {\bibfield  {journal} {\bibinfo
  {journal} {Phys. Rev. D}\ }\textbf {\bibinfo {volume} {91}},\ \bibinfo
  {pages} {034013} (\bibinfo {year} {2015})},\ \Eprint
  {http://arxiv.org/abs/1409.3205} {arXiv:1409.3205 [hep-ph]} \BibitemShut
  {NoStop}%
\bibitem [{\citenamefont {Appelquist}\ and\ \citenamefont
  {Politzer}(1975)}]{Appelquist:1974zd}%
  \BibitemOpen
  \bibfield  {author} {\bibinfo {author} {\bibfnamefont {T.}~\bibnamefont
  {Appelquist}}\ and\ \bibinfo {author} {\bibfnamefont {H.~D.}\ \bibnamefont
  {Politzer}},\ }\href {\doibase 10.1103/PhysRevLett.34.43} {\bibfield
  {journal} {\bibinfo  {journal} {Phys. Rev. Lett.}\ }\textbf {\bibinfo
  {volume} {34}},\ \bibinfo {pages} {43} (\bibinfo {year} {1975})}\BibitemShut
  {NoStop}%
\bibitem [{\citenamefont {Eichten}\ \emph {et~al.}(1978)\citenamefont
  {Eichten}, \citenamefont {Gottfried}, \citenamefont {Kinoshita},
  \citenamefont {Lane},\ and\ \citenamefont {Yan}}]{Eichten:1978tg}%
  \BibitemOpen
  \bibfield  {author} {\bibinfo {author} {\bibfnamefont {E.}~\bibnamefont
  {Eichten}}, \bibinfo {author} {\bibfnamefont {K.}~\bibnamefont {Gottfried}},
  \bibinfo {author} {\bibfnamefont {T.}~\bibnamefont {Kinoshita}}, \bibinfo
  {author} {\bibfnamefont {K.~D.}\ \bibnamefont {Lane}}, \ and\ \bibinfo
  {author} {\bibfnamefont {T.-M.}\ \bibnamefont {Yan}},\ }\href {\doibase
  10.1103/PhysRevD.17.3090} {\bibfield  {journal} {\bibinfo  {journal} {Phys.
  Rev. D}\ }\textbf {\bibinfo {volume} {17}},\ \bibinfo {pages} {3090}
  (\bibinfo {year} {1978})},\ \bibinfo {note} {[Erratum: Phys.Rev.D 21, 313
  (1980)]}\BibitemShut {NoStop}%
\bibitem [{\citenamefont {Godfrey}\ and\ \citenamefont
  {Isgur}(1985)}]{Godfrey:1985xj}%
  \BibitemOpen
  \bibfield  {author} {\bibinfo {author} {\bibfnamefont {S.}~\bibnamefont
  {Godfrey}}\ and\ \bibinfo {author} {\bibfnamefont {N.}~\bibnamefont
  {Isgur}},\ }\href {\doibase 10.1103/PhysRevD.32.189} {\bibfield  {journal}
  {\bibinfo  {journal} {Phys. Rev. D}\ }\textbf {\bibinfo {volume} {32}},\
  \bibinfo {pages} {189} (\bibinfo {year} {1985})}\BibitemShut {NoStop}%
\bibitem [{\citenamefont {Gupta}\ \emph {et~al.}(1994)\citenamefont {Gupta},
  \citenamefont {Johnson}, \citenamefont {Repko},\ and\ \citenamefont
  {Suchyta}}]{Gupta:1993pd}%
  \BibitemOpen
  \bibfield  {author} {\bibinfo {author} {\bibfnamefont {S.~N.}\ \bibnamefont
  {Gupta}}, \bibinfo {author} {\bibfnamefont {J.~M.}\ \bibnamefont {Johnson}},
  \bibinfo {author} {\bibfnamefont {W.~W.}\ \bibnamefont {Repko}}, \ and\
  \bibinfo {author} {\bibfnamefont {C.~J.}\ \bibnamefont {Suchyta},
  \bibfnamefont {III}},\ }\href {\doibase 10.1103/PhysRevD.49.1551} {\bibfield
  {journal} {\bibinfo  {journal} {Phys. Rev. D}\ }\textbf {\bibinfo {volume}
  {49}},\ \bibinfo {pages} {1551} (\bibinfo {year} {1994})},\ \Eprint
  {http://arxiv.org/abs/hep-ph/9312205} {arXiv:hep-ph/9312205} \BibitemShut
  {NoStop}%
\bibitem [{\citenamefont {Zeng}\ \emph {et~al.}(1995)\citenamefont {Zeng},
  \citenamefont {Van~Orden},\ and\ \citenamefont {Roberts}}]{Zeng:1994vj}%
  \BibitemOpen
  \bibfield  {author} {\bibinfo {author} {\bibfnamefont {J.}~\bibnamefont
  {Zeng}}, \bibinfo {author} {\bibfnamefont {J.~W.}\ \bibnamefont {Van~Orden}},
  \ and\ \bibinfo {author} {\bibfnamefont {W.}~\bibnamefont {Roberts}},\ }\href
  {\doibase 10.1103/PhysRevD.52.5229} {\bibfield  {journal} {\bibinfo
  {journal} {Phys. Rev. D}\ }\textbf {\bibinfo {volume} {52}},\ \bibinfo
  {pages} {5229} (\bibinfo {year} {1995})},\ \Eprint
  {http://arxiv.org/abs/hep-ph/9412269} {arXiv:hep-ph/9412269} \BibitemShut
  {NoStop}%
\bibitem [{\citenamefont {Ebert}\ \emph {et~al.}(2003)\citenamefont {Ebert},
  \citenamefont {Faustov},\ and\ \citenamefont {Galkin}}]{Ebert:2002pp}%
  \BibitemOpen
  \bibfield  {author} {\bibinfo {author} {\bibfnamefont {D.}~\bibnamefont
  {Ebert}}, \bibinfo {author} {\bibfnamefont {R.~N.}\ \bibnamefont {Faustov}},
  \ and\ \bibinfo {author} {\bibfnamefont {V.~O.}\ \bibnamefont {Galkin}},\
  }\href {\doibase 10.1103/PhysRevD.67.014027} {\bibfield  {journal} {\bibinfo
  {journal} {Phys. Rev. D}\ }\textbf {\bibinfo {volume} {67}},\ \bibinfo
  {pages} {014027} (\bibinfo {year} {2003})},\ \Eprint
  {http://arxiv.org/abs/hep-ph/0210381} {arXiv:hep-ph/0210381} \BibitemShut
  {NoStop}%
\bibitem [{\citenamefont {Bernardini}\ and\ \citenamefont
  {Dobrigkeit}(2003)}]{Bernardini:2003ht}%
  \BibitemOpen
  \bibfield  {author} {\bibinfo {author} {\bibfnamefont {A.~E.}\ \bibnamefont
  {Bernardini}}\ and\ \bibinfo {author} {\bibfnamefont {C.}~\bibnamefont
  {Dobrigkeit}},\ }\href {\doibase 10.1088/0954-3899/29/7/310} {\bibfield
  {journal} {\bibinfo  {journal} {J. Phys. G}\ }\textbf {\bibinfo {volume}
  {29}},\ \bibinfo {pages} {1439} (\bibinfo {year} {2003})},\ \Eprint
  {http://arxiv.org/abs/hep-ph/0611336} {arXiv:hep-ph/0611336} \BibitemShut
  {NoStop}%
\bibitem [{\citenamefont {Cucchieri}\ \emph {et~al.}(2019)\citenamefont
  {Cucchieri}, \citenamefont {Mendes},\ and\ \citenamefont
  {Serenone}}]{Cucchieri:2017icl}%
  \BibitemOpen
  \bibfield  {author} {\bibinfo {author} {\bibfnamefont {A.}~\bibnamefont
  {Cucchieri}}, \bibinfo {author} {\bibfnamefont {T.}~\bibnamefont {Mendes}}, \
  and\ \bibinfo {author} {\bibfnamefont {W.~M.}\ \bibnamefont {Serenone}},\
  }\href {\doibase 10.1007/s13538-019-00665-6} {\bibfield  {journal} {\bibinfo
  {journal} {Braz. J. Phys.}\ }\textbf {\bibinfo {volume} {49}},\ \bibinfo
  {pages} {548} (\bibinfo {year} {2019})},\ \Eprint
  {http://arxiv.org/abs/1704.08288} {arXiv:1704.08288 [hep-lat]} \BibitemShut
  {NoStop}%
\bibitem [{\citenamefont {Mutuk}(2018)}]{Mutuk:2018xay}%
  \BibitemOpen
  \bibfield  {author} {\bibinfo {author} {\bibfnamefont {H.}~\bibnamefont
  {Mutuk}},\ }\href {\doibase 10.1155/2018/5961031} {\bibfield  {journal}
  {\bibinfo  {journal} {Adv. High Energy Phys.}\ }\textbf {\bibinfo {volume}
  {2018}},\ \bibinfo {pages} {5961031} (\bibinfo {year} {2018})},\ \Eprint
  {http://arxiv.org/abs/1803.10603} {arXiv:1803.10603 [hep-ph]} \BibitemShut
  {NoStop}%
\bibitem [{\citenamefont {Gutierrez-Guerrero}\ \emph
  {et~al.}(2021)\citenamefont {Gutierrez-Guerrero}, \citenamefont {Alfaro},\
  and\ \citenamefont {Raya}}]{Gutierrez-Guerrero:2021fuj}%
  \BibitemOpen
  \bibfield  {author} {\bibinfo {author} {\bibfnamefont {L.~X.}\ \bibnamefont
  {Gutierrez-Guerrero}}, \bibinfo {author} {\bibfnamefont {J.}~\bibnamefont
  {Alfaro}}, \ and\ \bibinfo {author} {\bibfnamefont {A.}~\bibnamefont
  {Raya}},\ }\href {\doibase 10.1142/S0217751X21501712} {\bibfield  {journal}
  {\bibinfo  {journal} {Int. J. Mod. Phys. A}\ }\textbf {\bibinfo {volume}
  {36}},\ \bibinfo {pages} {2150171} (\bibinfo {year} {2021})},\ \Eprint
  {http://arxiv.org/abs/2108.12532} {arXiv:2108.12532 [hep-ph]} \BibitemShut
  {NoStop}%
\bibitem [{\citenamefont {Eichten}\ \emph {et~al.}(1980)\citenamefont
  {Eichten}, \citenamefont {Gottfried}, \citenamefont {Kinoshita},
  \citenamefont {Lane},\ and\ \citenamefont {Yan}}]{Eichten:1979ms}%
  \BibitemOpen
  \bibfield  {author} {\bibinfo {author} {\bibfnamefont {E.}~\bibnamefont
  {Eichten}}, \bibinfo {author} {\bibfnamefont {K.}~\bibnamefont {Gottfried}},
  \bibinfo {author} {\bibfnamefont {T.}~\bibnamefont {Kinoshita}}, \bibinfo
  {author} {\bibfnamefont {K.~D.}\ \bibnamefont {Lane}}, \ and\ \bibinfo
  {author} {\bibfnamefont {T.-M.}\ \bibnamefont {Yan}},\ }\href {\doibase
  10.1103/PhysRevD.21.203} {\bibfield  {journal} {\bibinfo  {journal} {Phys.
  Rev. D}\ }\textbf {\bibinfo {volume} {21}},\ \bibinfo {pages} {203} (\bibinfo
  {year} {1980})}\BibitemShut {NoStop}%
\bibitem [{\citenamefont {Eichten}\ \emph {et~al.}(2002)\citenamefont
  {Eichten}, \citenamefont {Lane},\ and\ \citenamefont
  {Quigg}}]{Eichten:2002qv}%
  \BibitemOpen
  \bibfield  {author} {\bibinfo {author} {\bibfnamefont {E.~J.}\ \bibnamefont
  {Eichten}}, \bibinfo {author} {\bibfnamefont {K.}~\bibnamefont {Lane}}, \
  and\ \bibinfo {author} {\bibfnamefont {C.}~\bibnamefont {Quigg}},\ }\href
  {\doibase 10.1103/PhysRevLett.89.162002} {\bibfield  {journal} {\bibinfo
  {journal} {Phys. Rev. Lett.}\ }\textbf {\bibinfo {volume} {89}},\ \bibinfo
  {pages} {162002} (\bibinfo {year} {2002})},\ \Eprint
  {http://arxiv.org/abs/hep-ph/0206018} {arXiv:hep-ph/0206018} \BibitemShut
  {NoStop}%
\bibitem [{\citenamefont {Eichten}\ and\ \citenamefont
  {Feinberg}(1981)}]{Eichten:1980mw}%
  \BibitemOpen
  \bibfield  {author} {\bibinfo {author} {\bibfnamefont {E.}~\bibnamefont
  {Eichten}}\ and\ \bibinfo {author} {\bibfnamefont {F.}~\bibnamefont
  {Feinberg}},\ }\href {\doibase 10.1103/PhysRevD.23.2724} {\bibfield
  {journal} {\bibinfo  {journal} {Phys. Rev. D}\ }\textbf {\bibinfo {volume}
  {23}},\ \bibinfo {pages} {2724} (\bibinfo {year} {1981})}\BibitemShut
  {NoStop}%
\bibitem [{\citenamefont {Gromes}(1984)}]{Gromes:1984ma}%
  \BibitemOpen
  \bibfield  {author} {\bibinfo {author} {\bibfnamefont {D.}~\bibnamefont
  {Gromes}},\ }\href {\doibase 10.1007/BF01452566} {\bibfield  {journal}
  {\bibinfo  {journal} {Z. Phys. C}\ }\textbf {\bibinfo {volume} {26}},\
  \bibinfo {pages} {401} (\bibinfo {year} {1984})}\BibitemShut {NoStop}%
\bibitem [{\citenamefont {Brambilla}\ \emph {et~al.}(2004)\citenamefont
  {Brambilla} \emph {et~al.}}]{QuarkoniumWorkingGroup:2004kpm}%
  \BibitemOpen
  \bibfield  {author} {\bibinfo {author} {\bibfnamefont {N.}~\bibnamefont
  {Brambilla}} \emph {et~al.} (\bibinfo {collaboration} {Quarkonium Working
  Group}),\ }\href {\doibase 10.5170/CERN-2005-005} {\  (\bibinfo {year}
  {2004}),\ 10.5170/CERN-2005-005},\ \Eprint
  {http://arxiv.org/abs/hep-ph/0412158} {arXiv:hep-ph/0412158} \BibitemShut
  {NoStop}%
\bibitem [{\citenamefont {Brambilla}\ \emph {et~al.}(2011)\citenamefont
  {Brambilla} \emph {et~al.}}]{Brambilla:2010cs}%
  \BibitemOpen
  \bibfield  {author} {\bibinfo {author} {\bibfnamefont {N.}~\bibnamefont
  {Brambilla}} \emph {et~al.},\ }\href {\doibase
  10.1140/epjc/s10052-010-1534-9} {\bibfield  {journal} {\bibinfo  {journal}
  {Eur. Phys. J. C}\ }\textbf {\bibinfo {volume} {71}},\ \bibinfo {pages}
  {1534} (\bibinfo {year} {2011})},\ \Eprint {http://arxiv.org/abs/1010.5827}
  {arXiv:1010.5827 [hep-ph]} \BibitemShut {NoStop}%
\bibitem [{\citenamefont {Parisi}\ and\ \citenamefont
  {Petronzio}(1980)}]{Parisi:1980jy}%
  \BibitemOpen
  \bibfield  {author} {\bibinfo {author} {\bibfnamefont {G.}~\bibnamefont
  {Parisi}}\ and\ \bibinfo {author} {\bibfnamefont {R.}~\bibnamefont
  {Petronzio}},\ }\href {\doibase 10.1016/0370-2693(80)90822-9} {\bibfield
  {journal} {\bibinfo  {journal} {Phys. Lett. B}\ }\textbf {\bibinfo {volume}
  {94}},\ \bibinfo {pages} {51} (\bibinfo {year} {1980})}\BibitemShut {NoStop}%
\bibitem [{\citenamefont {Consoli}\ and\ \citenamefont
  {Field}(1994)}]{Consoli:1993ew}%
  \BibitemOpen
  \bibfield  {author} {\bibinfo {author} {\bibfnamefont {M.}~\bibnamefont
  {Consoli}}\ and\ \bibinfo {author} {\bibfnamefont {J.~H.}\ \bibnamefont
  {Field}},\ }\href {\doibase 10.1103/PhysRevD.49.1293} {\bibfield  {journal}
  {\bibinfo  {journal} {Phys. Rev. D}\ }\textbf {\bibinfo {volume} {49}},\
  \bibinfo {pages} {1293} (\bibinfo {year} {1994})}\BibitemShut {NoStop}%
\bibitem [{\citenamefont {Consoli}\ and\ \citenamefont
  {Field}(1997)}]{Consoli:1997ts}%
  \BibitemOpen
  \bibfield  {author} {\bibinfo {author} {\bibfnamefont {M.}~\bibnamefont
  {Consoli}}\ and\ \bibinfo {author} {\bibfnamefont {J.~H.}\ \bibnamefont
  {Field}},\ }\href {\doibase 10.1088/0954-3899/23/1/004} {\bibfield  {journal}
  {\bibinfo  {journal} {J. Phys. G}\ }\textbf {\bibinfo {volume} {23}},\
  \bibinfo {pages} {41} (\bibinfo {year} {1997})}\BibitemShut {NoStop}%
\bibitem [{\citenamefont {Mihara}\ and\ \citenamefont
  {Natale}(2000)}]{Mihara:2000wf}%
  \BibitemOpen
  \bibfield  {author} {\bibinfo {author} {\bibfnamefont {A.}~\bibnamefont
  {Mihara}}\ and\ \bibinfo {author} {\bibfnamefont {A.~A.}\ \bibnamefont
  {Natale}},\ }\href {\doibase 10.1016/S0370-2693(00)00546-3} {\bibfield
  {journal} {\bibinfo  {journal} {Phys. Lett. B}\ }\textbf {\bibinfo {volume}
  {482}},\ \bibinfo {pages} {378} (\bibinfo {year} {2000})},\ \Eprint
  {http://arxiv.org/abs/hep-ph/0004236} {arXiv:hep-ph/0004236} \BibitemShut
  {NoStop}%
\bibitem [{\citenamefont {Field}(2002)}]{Field:2001iu}%
  \BibitemOpen
  \bibfield  {author} {\bibinfo {author} {\bibfnamefont {J.~H.}\ \bibnamefont
  {Field}},\ }\href {\doibase 10.1103/PhysRevD.66.013013} {\bibfield  {journal}
  {\bibinfo  {journal} {Phys. Rev. D}\ }\textbf {\bibinfo {volume} {66}},\
  \bibinfo {pages} {013013} (\bibinfo {year} {2002})},\ \Eprint
  {http://arxiv.org/abs/hep-ph/0101158} {arXiv:hep-ph/0101158} \BibitemShut
  {NoStop}%
\bibitem [{\citenamefont {Salpeter}\ and\ \citenamefont
  {Bethe}(1951)}]{Salpeter:1951sz}%
  \BibitemOpen
  \bibfield  {author} {\bibinfo {author} {\bibfnamefont {E.~E.}\ \bibnamefont
  {Salpeter}}\ and\ \bibinfo {author} {\bibfnamefont {H.~A.}\ \bibnamefont
  {Bethe}},\ }\href {\doibase 10.1103/PhysRev.84.1232} {\bibfield  {journal}
  {\bibinfo  {journal} {Phys. Rev.}\ }\textbf {\bibinfo {volume} {84}},\
  \bibinfo {pages} {1232} (\bibinfo {year} {1951})}\BibitemShut {NoStop}%
\bibitem [{\citenamefont {Bowman}\ \emph {et~al.}(2007)\citenamefont {Bowman},
  \citenamefont {Heller}, \citenamefont {Leinweber}, \citenamefont
  {Parappilly}, \citenamefont {Sternbeck}, \citenamefont {von Smekal},
  \citenamefont {Williams},\ and\ \citenamefont {Zhang}}]{Bowman:2007du}%
  \BibitemOpen
  \bibfield  {author} {\bibinfo {author} {\bibfnamefont {P.~O.}\ \bibnamefont
  {Bowman}}, \bibinfo {author} {\bibfnamefont {U.~M.}\ \bibnamefont {Heller}},
  \bibinfo {author} {\bibfnamefont {D.~B.}\ \bibnamefont {Leinweber}}, \bibinfo
  {author} {\bibfnamefont {M.~B.}\ \bibnamefont {Parappilly}}, \bibinfo
  {author} {\bibfnamefont {A.}~\bibnamefont {Sternbeck}}, \bibinfo {author}
  {\bibfnamefont {L.}~\bibnamefont {von Smekal}}, \bibinfo {author}
  {\bibfnamefont {A.~G.}\ \bibnamefont {Williams}}, \ and\ \bibinfo {author}
  {\bibfnamefont {J.-b.}\ \bibnamefont {Zhang}},\ }\href {\doibase
  10.1103/PhysRevD.76.094505} {\bibfield  {journal} {\bibinfo  {journal} {Phys.
  Rev. D}\ }\textbf {\bibinfo {volume} {76}},\ \bibinfo {pages} {094505}
  (\bibinfo {year} {2007})},\ \Eprint {http://arxiv.org/abs/hep-lat/0703022}
  {arXiv:hep-lat/0703022} \BibitemShut {NoStop}%
\bibitem [{\citenamefont {Oliveira}\ \emph {et~al.}(2019)\citenamefont
  {Oliveira}, \citenamefont {Silva}, \citenamefont {Skullerud},\ and\
  \citenamefont {Sternbeck}}]{Oliveira:2018lln}%
  \BibitemOpen
  \bibfield  {author} {\bibinfo {author} {\bibfnamefont {O.}~\bibnamefont
  {Oliveira}}, \bibinfo {author} {\bibfnamefont {P.~J.}\ \bibnamefont {Silva}},
  \bibinfo {author} {\bibfnamefont {J.-I.}\ \bibnamefont {Skullerud}}, \ and\
  \bibinfo {author} {\bibfnamefont {A.}~\bibnamefont {Sternbeck}},\ }\href
  {\doibase 10.1103/PhysRevD.99.094506} {\bibfield  {journal} {\bibinfo
  {journal} {Phys. Rev. D}\ }\textbf {\bibinfo {volume} {99}},\ \bibinfo
  {pages} {094506} (\bibinfo {year} {2019})},\ \Eprint
  {http://arxiv.org/abs/1809.02541} {arXiv:1809.02541 [hep-lat]} \BibitemShut
  {NoStop}%
\bibitem [{\citenamefont {Serreau}\ and\ \citenamefont
  {Tissier}(2012)}]{Serreau:2012cg}%
  \BibitemOpen
  \bibfield  {author} {\bibinfo {author} {\bibfnamefont {J.}~\bibnamefont
  {Serreau}}\ and\ \bibinfo {author} {\bibfnamefont {M.}~\bibnamefont
  {Tissier}},\ }\href {\doibase 10.1016/j.physletb.2012.04.041} {\bibfield
  {journal} {\bibinfo  {journal} {Phys. Lett. B}\ }\textbf {\bibinfo {volume}
  {712}},\ \bibinfo {pages} {97} (\bibinfo {year} {2012})},\ \Eprint
  {http://arxiv.org/abs/1202.3432} {arXiv:1202.3432 [hep-th]} \BibitemShut
  {NoStop}%
\bibitem [{\citenamefont {Lucha}\ and\ \citenamefont
  {Schoberl}(1995)}]{Lucha:1995zv}%
  \BibitemOpen
  \bibfield  {author} {\bibinfo {author} {\bibfnamefont {W.}~\bibnamefont
  {Lucha}}\ and\ \bibinfo {author} {\bibfnamefont {F.~F.}\ \bibnamefont
  {Schoberl}},\ }in\ \href@noop {} {\emph {\bibinfo {booktitle} {{International
  Summer School for Students on Development in Nuclear Theory and Particle
  Physics}}}}\ (\bibinfo {year} {1995})\ \Eprint
  {http://arxiv.org/abs/hep-ph/9601263} {arXiv:hep-ph/9601263} \BibitemShut
  {NoStop}%
\bibitem [{\citenamefont {Caswell}\ and\ \citenamefont
  {Lepage}(1978)}]{Caswell:1978mt}%
  \BibitemOpen
  \bibfield  {author} {\bibinfo {author} {\bibfnamefont {W.~E.}\ \bibnamefont
  {Caswell}}\ and\ \bibinfo {author} {\bibfnamefont {G.~P.}\ \bibnamefont
  {Lepage}},\ }\href {\doibase 10.1103/PhysRevA.18.810} {\bibfield  {journal}
  {\bibinfo  {journal} {Phys. Rev. A}\ }\textbf {\bibinfo {volume} {18}},\
  \bibinfo {pages} {810} (\bibinfo {year} {1978})}\BibitemShut {NoStop}%
\bibitem [{\citenamefont {Gromes}(1977)}]{Gromes:1976np}%
  \BibitemOpen
  \bibfield  {author} {\bibinfo {author} {\bibfnamefont {D.}~\bibnamefont
  {Gromes}},\ }\href {\doibase 10.1016/0550-3213(77)90186-9} {\bibfield
  {journal} {\bibinfo  {journal} {Nucl. Phys. B}\ }\textbf {\bibinfo {volume}
  {131}},\ \bibinfo {pages} {80} (\bibinfo {year} {1977})}\BibitemShut
  {NoStop}%
\bibitem [{\citenamefont {Breit}(1929)}]{Breit1}%
  \BibitemOpen
  \bibfield  {author} {\bibinfo {author} {\bibfnamefont {G.}~\bibnamefont
  {Breit}},\ }\href {\doibase 10.1103/PhysRev.34.553} {\bibfield  {journal}
  {\bibinfo  {journal} {Phys. Rev.}\ }\textbf {\bibinfo {volume} {34}},\
  \bibinfo {pages} {553} (\bibinfo {year} {1929})}\BibitemShut {NoStop}%
\bibitem [{\citenamefont {Breit}(1930)}]{Breit2}%
  \BibitemOpen
  \bibfield  {author} {\bibinfo {author} {\bibfnamefont {G.}~\bibnamefont
  {Breit}},\ }\href {\doibase 10.1103/PhysRev.36.383} {\bibfield  {journal}
  {\bibinfo  {journal} {Phys. Rev.}\ }\textbf {\bibinfo {volume} {36}},\
  \bibinfo {pages} {383} (\bibinfo {year} {1930})}\BibitemShut {NoStop}%
\bibitem [{\citenamefont {Breit}(1932)}]{Breit3}%
  \BibitemOpen
  \bibfield  {author} {\bibinfo {author} {\bibfnamefont {G.}~\bibnamefont
  {Breit}},\ }\href {\doibase 10.1103/PhysRev.39.616} {\bibfield  {journal}
  {\bibinfo  {journal} {Phys. Rev.}\ }\textbf {\bibinfo {volume} {39}},\
  \bibinfo {pages} {616} (\bibinfo {year} {1932})}\BibitemShut {NoStop}%
\bibitem [{\citenamefont {Eiglsperger}(2007)}]{Eiglsperger:2007ay}%
  \BibitemOpen
  \bibfield  {author} {\bibinfo {author} {\bibfnamefont {J.}~\bibnamefont
  {Eiglsperger}},\ }\emph {\bibinfo {title} {{Quarkonium Spectroscopy: Beyond
  One-Gluon Exchange}}},\ \href@noop {} {Master's thesis},\ \bibinfo  {school}
  {Munich, Tech. U.} (\bibinfo {year} {2007}),\ \Eprint
  {http://arxiv.org/abs/0707.1269} {arXiv:0707.1269 [hep-ph]} \BibitemShut
  {NoStop}%
\bibitem [{\citenamefont {Eichten}\ \emph {et~al.}(2008)\citenamefont
  {Eichten}, \citenamefont {Godfrey}, \citenamefont {Mahlke},\ and\
  \citenamefont {Rosner}}]{Eichten:2007qx}%
  \BibitemOpen
  \bibfield  {author} {\bibinfo {author} {\bibfnamefont {E.}~\bibnamefont
  {Eichten}}, \bibinfo {author} {\bibfnamefont {S.}~\bibnamefont {Godfrey}},
  \bibinfo {author} {\bibfnamefont {H.}~\bibnamefont {Mahlke}}, \ and\ \bibinfo
  {author} {\bibfnamefont {J.~L.}\ \bibnamefont {Rosner}},\ }\href {\doibase
  10.1103/RevModPhys.80.1161} {\bibfield  {journal} {\bibinfo  {journal} {Rev.
  Mod. Phys.}\ }\textbf {\bibinfo {volume} {80}},\ \bibinfo {pages} {1161}
  (\bibinfo {year} {2008})},\ \Eprint {http://arxiv.org/abs/hep-ph/0701208}
  {arXiv:hep-ph/0701208} \BibitemShut {NoStop}%
\bibitem [{\citenamefont {Lucha}\ \emph {et~al.}(1991)\citenamefont {Lucha},
  \citenamefont {Schoberl},\ and\ \citenamefont {Gromes}}]{Lucha:1991vn}%
  \BibitemOpen
  \bibfield  {author} {\bibinfo {author} {\bibfnamefont {W.}~\bibnamefont
  {Lucha}}, \bibinfo {author} {\bibfnamefont {F.~F.}\ \bibnamefont {Schoberl}},
  \ and\ \bibinfo {author} {\bibfnamefont {D.}~\bibnamefont {Gromes}},\ }\href
  {\doibase 10.1016/0370-1573(91)90001-3} {\bibfield  {journal} {\bibinfo
  {journal} {Phys. Rept.}\ }\textbf {\bibinfo {volume} {200}},\ \bibinfo
  {pages} {127} (\bibinfo {year} {1991})}\BibitemShut {NoStop}%
\bibitem [{\citenamefont {Cahn}\ and\ \citenamefont
  {Jackson}(2003)}]{Cahn:2003cw}%
  \BibitemOpen
  \bibfield  {author} {\bibinfo {author} {\bibfnamefont {R.~N.}\ \bibnamefont
  {Cahn}}\ and\ \bibinfo {author} {\bibfnamefont {J.~D.}\ \bibnamefont
  {Jackson}},\ }\href {\doibase 10.1103/PhysRevD.68.037502} {\bibfield
  {journal} {\bibinfo  {journal} {Phys. Rev. D}\ }\textbf {\bibinfo {volume}
  {68}},\ \bibinfo {pages} {037502} (\bibinfo {year} {2003})},\ \Eprint
  {http://arxiv.org/abs/hep-ph/0305012} {arXiv:hep-ph/0305012} \BibitemShut
  {NoStop}%
\bibitem [{\citenamefont {Navas}\ \emph {et~al.}(2024)\citenamefont {Navas}
  \emph {et~al.}}]{ParticleDataGroup:2024cfk}%
  \BibitemOpen
  \bibfield  {author} {\bibinfo {author} {\bibfnamefont {S.}~\bibnamefont
  {Navas}} \emph {et~al.} (\bibinfo {collaboration} {Particle Data Group}),\
  }\href {\doibase 10.1103/PhysRevD.110.030001} {\bibfield  {journal} {\bibinfo
   {journal} {Phys. Rev. D}\ }\textbf {\bibinfo {volume} {110}},\ \bibinfo
  {pages} {030001} (\bibinfo {year} {2024})}\BibitemShut {NoStop}%
\bibitem [{\citenamefont {Mathur}\ \emph {et~al.}(2018)\citenamefont {Mathur},
  \citenamefont {Padmanath},\ and\ \citenamefont {Mondal}}]{Mathur:2018epb}%
  \BibitemOpen
  \bibfield  {author} {\bibinfo {author} {\bibfnamefont {N.}~\bibnamefont
  {Mathur}}, \bibinfo {author} {\bibfnamefont {M.}~\bibnamefont {Padmanath}}, \
  and\ \bibinfo {author} {\bibfnamefont {S.}~\bibnamefont {Mondal}},\ }\href
  {\doibase 10.1103/PhysRevLett.121.202002} {\bibfield  {journal} {\bibinfo
  {journal} {Phys. Rev. Lett.}\ }\textbf {\bibinfo {volume} {121}},\ \bibinfo
  {pages} {202002} (\bibinfo {year} {2018})},\ \Eprint
  {http://arxiv.org/abs/1806.04151} {arXiv:1806.04151 [hep-lat]} \BibitemShut
  {NoStop}%
\bibitem [{\citenamefont {Deur}\ \emph {et~al.}(2016)\citenamefont {Deur},
  \citenamefont {Brodsky},\ and\ \citenamefont {de~Teramond}}]{Deur:2016tte}%
  \BibitemOpen
  \bibfield  {author} {\bibinfo {author} {\bibfnamefont {A.}~\bibnamefont
  {Deur}}, \bibinfo {author} {\bibfnamefont {S.~J.}\ \bibnamefont {Brodsky}}, \
  and\ \bibinfo {author} {\bibfnamefont {G.~F.}\ \bibnamefont {de~Teramond}},\
  }\href {\doibase 10.1016/j.ppnp.2016.04.003} {\bibfield  {journal} {\bibinfo
  {journal} {Nucl. Phys.}\ }\textbf {\bibinfo {volume} {90}},\ \bibinfo {pages}
  {1} (\bibinfo {year} {2016})},\ \Eprint {http://arxiv.org/abs/1604.08082}
  {arXiv:1604.08082 [hep-ph]} \BibitemShut {NoStop}%
\bibitem [{\citenamefont {Ding}\ \emph {et~al.}(1995)\citenamefont {Ding},
  \citenamefont {Chao},\ and\ \citenamefont {Qin}}]{Ding:1995he}%
  \BibitemOpen
  \bibfield  {author} {\bibinfo {author} {\bibfnamefont {Y.-B.}\ \bibnamefont
  {Ding}}, \bibinfo {author} {\bibfnamefont {K.-T.}\ \bibnamefont {Chao}}, \
  and\ \bibinfo {author} {\bibfnamefont {D.-H.}\ \bibnamefont {Qin}},\ }\href
  {\doibase 10.1103/PhysRevD.51.5064} {\bibfield  {journal} {\bibinfo
  {journal} {Phys. Rev. D}\ }\textbf {\bibinfo {volume} {51}},\ \bibinfo
  {pages} {5064} (\bibinfo {year} {1995})},\ \Eprint
  {http://arxiv.org/abs/hep-ph/9502409} {arXiv:hep-ph/9502409} \BibitemShut
  {NoStop}%
\bibitem [{\citenamefont {Domenech-Garret}\ and\ \citenamefont
  {Sanchis-Lozano}(2009)}]{Domenech-Garret:2008itl}%
  \BibitemOpen
  \bibfield  {author} {\bibinfo {author} {\bibfnamefont {J.-L.}\ \bibnamefont
  {Domenech-Garret}}\ and\ \bibinfo {author} {\bibfnamefont {M.-A.}\
  \bibnamefont {Sanchis-Lozano}},\ }\href {\doibase 10.1016/j.cpc.2008.11.011}
  {\bibfield  {journal} {\bibinfo  {journal} {Comput. Phys. Commun.}\ }\textbf
  {\bibinfo {volume} {180}},\ \bibinfo {pages} {768} (\bibinfo {year}
  {2009})},\ \Eprint {http://arxiv.org/abs/0805.2704} {arXiv:0805.2704
  [hep-ph]} \BibitemShut {NoStop}%
\bibitem [{\citenamefont {Debastiani}\ and\ \citenamefont
  {Navarra}(2019)}]{Debastiani:2017msn}%
  \BibitemOpen
  \bibfield  {author} {\bibinfo {author} {\bibfnamefont {V.~R.}\ \bibnamefont
  {Debastiani}}\ and\ \bibinfo {author} {\bibfnamefont {F.~S.}\ \bibnamefont
  {Navarra}},\ }\href {\doibase 10.1088/1674-1137/43/1/013105} {\bibfield
  {journal} {\bibinfo  {journal} {Chin. Phys. C}\ }\textbf {\bibinfo {volume}
  {43}},\ \bibinfo {pages} {013105} (\bibinfo {year} {2019})},\ \Eprint
  {http://arxiv.org/abs/1706.07553} {arXiv:1706.07553 [hep-ph]} \BibitemShut
  {NoStop}%
\bibitem [{\citenamefont {Fagundes}\ \emph {et~al.}(2012)\citenamefont
  {Fagundes}, \citenamefont {Luna}, \citenamefont {Menon},\ and\ \citenamefont
  {Natale}}]{Fagundes:2011zx}%
  \BibitemOpen
  \bibfield  {author} {\bibinfo {author} {\bibfnamefont {D.~A.}\ \bibnamefont
  {Fagundes}}, \bibinfo {author} {\bibfnamefont {E.~G.~S.}\ \bibnamefont
  {Luna}}, \bibinfo {author} {\bibfnamefont {M.~J.}\ \bibnamefont {Menon}}, \
  and\ \bibinfo {author} {\bibfnamefont {A.~A.}\ \bibnamefont {Natale}},\
  }\href {\doibase 10.1016/j.nuclphysa.2012.05.002} {\bibfield  {journal}
  {\bibinfo  {journal} {Nucl. Phys. A}\ }\textbf {\bibinfo {volume} {886}},\
  \bibinfo {pages} {48} (\bibinfo {year} {2012})},\ \Eprint
  {http://arxiv.org/abs/1112.4680} {arXiv:1112.4680 [hep-ph]} \BibitemShut
  {NoStop}%
\bibitem [{\citenamefont {Guti\'errez-Guerrero}\ \emph
  {et~al.}(2019)\citenamefont {Guti\'errez-Guerrero}, \citenamefont {Bashir},
  \citenamefont {Bedolla},\ and\ \citenamefont
  {Santopinto}}]{Gutierrez-Guerrero:2019uwa}%
  \BibitemOpen
  \bibfield  {author} {\bibinfo {author} {\bibfnamefont {L.~X.}\ \bibnamefont
  {Guti\'errez-Guerrero}}, \bibinfo {author} {\bibfnamefont {A.}~\bibnamefont
  {Bashir}}, \bibinfo {author} {\bibfnamefont {M.~A.}\ \bibnamefont {Bedolla}},
  \ and\ \bibinfo {author} {\bibfnamefont {E.}~\bibnamefont {Santopinto}},\
  }\href {\doibase 10.1103/PhysRevD.100.114032} {\bibfield  {journal} {\bibinfo
   {journal} {Phys. Rev. D}\ }\textbf {\bibinfo {volume} {100}},\ \bibinfo
  {pages} {114032} (\bibinfo {year} {2019})},\ \Eprint
  {http://arxiv.org/abs/1911.09213} {arXiv:1911.09213 [nucl-th]} \BibitemShut
  {NoStop}%
\bibitem [{\citenamefont {Paredes-Torres}\ \emph {et~al.}(2024)\citenamefont
  {Paredes-Torres}, \citenamefont {Guti\'errez-Guerrero}, \citenamefont
  {Bashir},\ and\ \citenamefont {Miramontes}}]{Paredes-Torres:2024mnz}%
  \BibitemOpen
  \bibfield  {author} {\bibinfo {author} {\bibfnamefont {G.}~\bibnamefont
  {Paredes-Torres}}, \bibinfo {author} {\bibfnamefont {L.~X.}\ \bibnamefont
  {Guti\'errez-Guerrero}}, \bibinfo {author} {\bibfnamefont {A.}~\bibnamefont
  {Bashir}}, \ and\ \bibinfo {author} {\bibfnamefont {A.~S.}\ \bibnamefont
  {Miramontes}},\ }\href {\doibase 10.1103/PhysRevD.109.114006} {\bibfield
  {journal} {\bibinfo  {journal} {Phys. Rev. D}\ }\textbf {\bibinfo {volume}
  {109}},\ \bibinfo {pages} {114006} (\bibinfo {year} {2024})},\ \Eprint
  {http://arxiv.org/abs/2405.06101} {arXiv:2405.06101 [hep-ph]} \BibitemShut
  {NoStop}%
\bibitem [{\citenamefont {Guti\'errez-Guerrero}\ \emph
  {et~al.}(2024)\citenamefont {Guti\'errez-Guerrero}, \citenamefont {Raya},
  \citenamefont {Albino},\ and\ \citenamefont
  {Hern\'andez-Pinto}}]{Gutierrez-Guerrero:2024him}%
  \BibitemOpen
  \bibfield  {author} {\bibinfo {author} {\bibfnamefont {L.~X.}\ \bibnamefont
  {Guti\'errez-Guerrero}}, \bibinfo {author} {\bibfnamefont {A.}~\bibnamefont
  {Raya}}, \bibinfo {author} {\bibfnamefont {L.}~\bibnamefont {Albino}}, \ and\
  \bibinfo {author} {\bibfnamefont {R.~J.}\ \bibnamefont {Hern\'andez-Pinto}},\
  }\href {\doibase 10.1103/PhysRevD.110.074015} {\bibfield  {journal} {\bibinfo
   {journal} {Phys. Rev. D}\ }\textbf {\bibinfo {volume} {110}},\ \bibinfo
  {pages} {074015} (\bibinfo {year} {2024})},\ \Eprint
  {http://arxiv.org/abs/2409.06057} {arXiv:2409.06057 [hep-ph]} \BibitemShut
  {NoStop}%
\end{thebibliography}%

\end{document}